\documentclass[11pt,a4paper]{article}
\pdfoutput=1

\usepackage[utf8]{inputenc}
\usepackage{a4wide}
\usepackage{amsmath}
\usepackage{cite}
\usepackage[table]{xcolor}
\usepackage{amssymb}
\usepackage{amsfonts}

\usepackage{booktabs}
\usepackage{makecell}
\usepackage{multirow}
\usepackage{pdflscape}
\usepackage{textcomp}
\usepackage{gensymb}
\usepackage{bigstrut}
\usepackage{array}
\usepackage{enumerate}
\usepackage[small,bf]{caption}
\usepackage{subcaption}
\usepackage{diagbox}
\usepackage{graphicx}
\usepackage{hyperref}
\usepackage{appendix}
\usepackage{verbatim}
\usepackage{geometry}
\usepackage{listings}
\usepackage{mathtools}
\usepackage{longtable}
\usepackage{threeparttable}
\usepackage{floatpag}
\usepackage{framed}
\usepackage{calrsfs}
\usepackage[normalem]{ulem}
\DeclareMathAlphabet{\pazocal}{OMS}{zplm}{m}{n}
\SetMathAlphabet\pazocal{bold}{OMS}{zplm}{bx}{n}
\usepackage{cleveref} 
\crefname{figure}{Fig.}{Figs.}

\usepackage{xspace}

\usepackage{multicol}
\usepackage{afterpage}

\makeatletter
\newcommand {\crefext}[2]{\csname cref@#1@format\endcsname{#2}{}{}}
\newcommand {\Crefext}[2]{\csname Cref@#1@format\endcsname{#2}{}{}}
\makeatother

\DeclareMathOperator{\erf}{erf}

\newcommand\parbar{\smash[b]{
\raisebox{0.3\height}{\scalebox{0.5}{\tiny(}}
{\mkern-1.5mu
\scriptscriptstyle-
\mkern-1.5mu}
\raisebox{0.3\height}{\scalebox{0.5}{\tiny)}}
}}
\makeatletter
\newcommand{\oset}[3][0ex]{%
  \mathrel{\mathop{#3}\limits^{
    \vbox to#1{\kern-2\ex@
    \hbox{$\scriptstyle#2$}\vss}}}}
\makeatother

\newcommand\nuebp{{\oset{\parbar}{\nu}_{\hspace{-.45ex}e}}}
\newcommand\nuxbp{{\oset{\parbar}{\nu}_{\hspace{-.45ex}x}}}

\newcommand\A{\textbf{A}\xspace}
\newcommand\B{\textbf{B}\xspace}

\allowdisplaybreaks
\numberwithin{equation}{section}



\begin{document}

%%%%%%%%%%%%%%%%%%%%%%%
\begin{titlepage}

\begin{center}
{\bf\LARGE Supernova $\nu$ flavour conversions in DUNE: \\[2mm] the slow, the fast and the standard
}\\[8mm]
A.~Giarnetti\(^{\,a}\)\footnote{E-mail: \texttt{agiarnetti@roma1.infn.it}},
J.~T.~Penedo\(^{\,b,}\)\footnote{E-mail: \texttt{jpenedo@roma3.infn.it}}\\
 \vspace{5mm}
\(^{a}\)\,{\it INFN Sezione di Roma, Piazzale Aldo Moro 2, 00185,
Roma, Italy} \\
\vspace{2mm}
\(^{b}\)\,{\it INFN Sezione di Roma Tre, Via della Vasca Navale 84, 00146, Roma, Italy} 
\end{center}
\vspace{2mm}

\begin{abstract}
The flavour composition of a future supernova neutrino signal is expected to carry measurable imprints of flavour conversion processes in the dense stellar medium.
In this work, we analyse the sensitivity of the upcoming Deep Underground Neutrino Experiment (DUNE) to three phenomenologically distinct effects: slow energy-dependent collective oscillations, fast energy-independent collective oscillations, and standard MSW conversions.
By integrating GLoBES and MultiNest and using benchmark neutrino fluxes at emission,
we assess the potential of DUNE to extract the underlying flux parameters and discriminate among conversion scenarios.
\end{abstract}

\end{titlepage}
\setcounter{footnote}{0}
%%%%%%%%%%%%%%%%%%%%%%%

%%%%%%%%%%%%%%%%%%%%%%%%%%%%%%%%%%%%%%%%%%%%%%
\section{Introduction}
\label{sec:intro}
%%%%%%%%%%%%%%%%%%%%%%%%%%%%%%%%%%%%%%%%%%%%%%

Core-collapse supernovae (SNe) can be considered extreme neutrino factories,
producing an unparalleled number of these particles within a short interval of time.
In our Galaxy, they are estimated to occur at a rate of about one or two per century~\cite{Rozwadowska:2020nab}.
The study of the properties of SN neutrinos
as they reach the Earth, via their detection in terrestrial experiments,
could provide significant
insights into
neutrino physics in the extreme supernova conditions, as well as
into the processes involved in astrophysical phenomena~\cite{Raffelt:2007nv,Janka:2017vlw,Janka:2017vcp,Horiuchi:2018ofe,Volpe:2023met}.
Following the advent of large-scale neutrino detectors,
only one supernova, SN 1987A,
was close enough to produce
an observable neutrino flux on Earth,
a pivotal milestone of early neutrino astronomy \cite{Hirata:1988ad}.
Since then, the study of SN neutrinos
has attracted a growing interest.
In particular, several studies were performed to investigate the SN explosion mechanisms
and the expected neutrino spectrum generated by the catastrophic event, exploiting
different proposals for neutrino detection and the interpretation
of data, see e.g.~\cite{Loredo:2001rx,Keil:2002in,Pagliaroli:2008ur,Hudepohl:2009tyy,Tamborra:2012ac,GalloRosso:2017hbp,GalloRosso:2017mdz,GalloRosso:2018ugl,Shalgar:2019rqe,Hyper-Kamiokande:2021frf,Vissani:2021jxf,Olsen:2022pkn,Suwa:2022vhp,deGouvea:2022dtw,Harada:2023elm,Fiorillo:2023ytr,Fiorillo:2023cas,Saez:2024ayk,Abbar:2024nhz,Sen:2024fxa}.

The determination of the neutrino energy spectrum at the source is not the only
challenge for the interpretation of a future SN neutrino signal. Indeed,
the knowledge of the (energy-dependent) flavour breakdown of the SN neutrino flux as it reaches the surface of the Earth is of crucial importance.
The well-established phenomenon of neutrino oscillations~\cite{Bilenky:1978nj,Super-Kamiokande:1998kpq,SNO:2002tuh,DayaBay:2012fng,Kajita:2016cak,McDonald:2016ixn}
allows the neutrinos and antineutrinos produced by the explosion%
\footnote{Notice that in the first phase of the explosion, namely the neutronization burst, the emitted neutrinos are predominantly $\nu_e$, while in the following accretion and cooling phases all the neutrino and antineutrino flavours are emitted with comparable luminosities.
} 
to change their flavour while travelling to the detector.
However, apart from standard vacuum propagation and the Mikheyev–Smirnov–Wolfenstein (MSW) effect in matter~\cite{Wolfenstein:1977ue,Mikheyev:1985zog,Dighe:1999bi},
the extreme density of neutrinos themselves, as they start to free-stream, can trigger
so-called collective neutrino oscillations~\cite{Pantaleone:1992eq,Sigl:1993ctk,Pastor:2002we,Duan:2007mv,Dasgupta:2007ws,Mirizzi:2015eza,Chakraborty:2016yeg,Tamborra:2020cul,Dasgupta:2021gfs}
due to neutrino-neutrino forward scattering.
Unlike standard oscillation regimes, in which the neutrino flavour evolution is linear and the computation of oscillation probabilities is straightforward, in this case the dense-medium flavour Hamiltonian contains self-interactions that depend on the density matrix itself. This renders the dynamics non-linear and leads to an extremely rich phenomenology.

Two striking phenomena that result from collective effects and potentially affect a prospective signal
are fast flavour conversions (FFCs) and spectral swaps arising from slow flavour instabilities.
FFCs arise when the neutrino flavour evolution is dominated by
the self-interaction term and can occur on sub-microsecond timescales.
In particular,
if the angular distributions of $\nu_e$ and $\bar\nu_e$ feature a crossing, 
a fast instability is triggered and can lead to partial flavour equilibration~\cite{Sawyer:2005jk,Sawyer:2015dsa,Dasgupta:2016dbv,Izaguirre:2016gsx,Abbar:2018shq,Yi:2019hrp,Johns:2019izj,Bhattacharyya:2020dhu,Bhattacharyya:2020jpj,Bhattacharyya:2021klg,Martin:2021xyl,Morinaga:2021vmc,Wu:2021uvt,Bhattacharyya:2022eed,Richers:2022zug,Zaizen:2022cik,Ehring:2023lcd,Xiong:2023vcm,Nagakura:2023xhc,Abbar:2023ltx,Abbar:2024ynh,Xiong:2024tac,Fiorillo:2024bzm,Fiorillo:2024uki,Fiorillo:2025npi}.
Instead, the interference of the vacuum ($\sim\Delta m^2/E$) and
self-interaction ($\sim G_F\, n_\nu$) terms
is responsible for slow instabilities,
which develop over tens of milliseconds.
These occur in the presence of spectral crossings
and can induce distinctive spectral swapping~\cite{Duan:2006an,Duan:2006jv,Hannestad:2006nj,Raffelt:2007cb,Duan:2007bt,Fogli:2007bk,Raffelt:2007xt,Dasgupta:2008cd,Dasgupta:2009mg,Fogli:2009rd,Duan:2010bg,Galais:2011gh},
i.e.~the interchange of the $\nu_e$ and $\nu_{\mu,\tau}$ spectra
(or $\bar\nu_e \leftrightarrow \bar\nu_{\mu,\tau}$) over a range of energies.
Note that, despite their names, slow instabilities can substantially develop
before fast ones in the supernova environment~\cite{Fiorillo:2025gkw}.

\vskip 2mm

The complexity intrinsic to the treatment of collective effects has led to the development of two separate but complementary approaches in SN neutrino studies.
One approach focuses on the simulation of the SN evolution, including neutrino production and transport. While the prevalence of FFCs has been recognized, these studies still face the challenge of consistently including them in radiation-hydrodynamics simulations, in part due to the range of scales involved (see~\cite{Johns:2025mlm} for a recent review).
Nevertheless, several analytical prescriptions
for quantifying the impact of FFCs on the neutrino flux have recently been put forward in the literature~\cite{Bhattacharyya:2020dhu,Bhattacharyya:2020jpj,Wu:2021uvt,Zaizen:2022cik,Bhattacharyya:2022eed,Xiong:2023vcm,George:2024zxz} (see also~\cite{Capanema:2024hdm}).
The other approach centres on neutrino detection at future facilities, aiming to assess how a given detector may constrain neutrino properties (e.g.~the mass ordering) or discriminate among SN models~\cite{Choubey:2010up,Seadrow:2018ftp,Nagakura:2020qhb,DUNE:2020zfm,Nagakura:2021lma,DUNE:2023rtr,Panda:2023rxa,Huang:2023aob,Hajjar:2023xae,Saez:2023snv,Gaba:2024asp,Choi:2025igp}.
Such analyses typically rely on post-processed neutrino fluxes from simulations and often overlook collective effects, with FFCs in particular being either omitted or assumed to lead to complete flavour equilibration.

This work aims at expanding the tentative bridge connecting these two worlds.
Taking a phenomenological approach, we parameterize the neutrino spectra and angular distributions and implement analytical prescriptions to fold in the impact of slow and fast instabilities and standard conversions.
This allows for the computation of the expected neutrino and antineutrino event rates at the Earth and to compare flavour conversion scenarios. In this context, a Bayesian comparison has previously been carried out in Ref.~\cite{Abbar:2024nhz} for water Cherenkov detectors.
In this work, we concentrate on the supernova-neutrino detection capabilities of the upcoming Deep Underground Neutrino Experiment (DUNE)~\cite{DUNE:2015lol} Far Detector (FD)~\cite{DUNE:2020ypp},
a liquid argon time-projection chamber expected to record $\pazocal{O}(10^3)$ events from a future SN at 10 kpc~\cite{DUNE:2020zfm,DUNE:2023rtr}.
By making use of the General Long Baseline Experiment Simulator (GLoBES) and integrating it with the Bayesian inference tool MultiNest, we are able to estimate the expected sensitivity of the experiment to the values of the parameters describing the neutrino flux.%
\footnote{The flux parameter estimations in\mbox{\cite{DUNE:2020zfm,DUNE:2023rtr,Huang:2023aob}}
 were performed in the absence of collective effects.
}

The paper is organized as follows.
We introduce our setup in~\cref{sec:SNnus}, including the analytical approximations and parameterizations relevant to the estimation of the effects of flavour conversions, as well as the chosen benchmarks.
In~\cref{sec:DUNE}, we describe the different channels for neutrino detection and examine the time-integrated neutrino event rates at the DUNE FD, in the absence of flavour conversions.
Then, in~\cref{sec:results}, 
we illustrate the effect of different conversion scenarios on the neutrino fluences and further study the extent to which the data would allow to estimate flux parameters and distinguish conversion scenarios, in the presence or absence of fast and slow collective effects, as well as standard MSW conversions (for both mass orderings).
We summarize our results in~\cref{sec:summary}.

%%%%%%%%%%%%%%%%%%%%%%%%%%%%%%%%%%%%%%%%%%%%%%
\section{Supernova neutrinos}
\label{sec:SNnus}
%%%%%%%%%%%%%%%%%%%%%%%%%%%%%%%%%%%%%%%%%%%%%%

\subsection{Fluxes, fluences and luminosities}

The time-dependent spectral flux $d\dot{\Phi}_i/dE$ of the neutrino (or antineutrino) species $i$ at an Earth-based detector is given by
\begin{equation}
\frac{d\dot{\Phi}_i}{dE} = \frac{1}{4\pi D^2} \frac{d\dot{\mathsf{N}}_i}{dE} \,,
\end{equation}
where $\dot{\mathsf{N}}_i$ is the emission rate at the neutrinosphere and $D$ is the distance to the supernova. The spectral fluence $d\Phi_i/dE$ at the Earth is simply the time-integrated (spectral) flux,
\begin{equation}
\frac{d\Phi_i}{dE} = \int_{\Delta T} dt\, \frac{d\dot{\Phi}_i}{dE} \,,
\end{equation}
and has dimensions of an inverse energy times an inverse area. The total number of events at one detector, $dN_i/dE$, at a given energy and over a time period $\Delta T$ will then be proportional to -- or result from the convolution with -- the corresponding fluence. 

The luminosity (or emitted power) associated with species $i$, in units of energy over time, instead reads
\begin{equation} \label{eq:luminosity}
\mathcal{L}_i=\int_0^\infty \frac{d\mathcal{L}_i}{dE}\, dE 
 =\int_0^\infty \frac{d\dot{\mathsf{N}}_i}{dE}\, E\, dE 
\,.
\end{equation}
The spectrum of radiated neutrinos can further be parameterized using a ``pinched'' form
\begin{equation} \label{eq:pinched}
\frac{d\dot{\mathsf{N}}_i}{dE} \,=\, \frac{1}{E} \frac{d\mathcal{L}_i}{dE}
\,=\, \frac{\mathcal{L}_i}{\langle E_i \rangle^2} \frac{(\alpha_i +1)^{\alpha_i +1}}{\Gamma(\alpha_i +1)} \left(\frac{E}{\langle E_i\rangle}\right)^{\alpha_i}\, e^{-(\alpha_i +1) \frac{E}{\langle E_i \rangle}}\,,
\end{equation}
with reasonably good accuracy~\cite{Keil:2002in,Tamborra:2012ac}.
Here, the $\langle E_i \rangle$ are the average neutrino energies, while the $\alpha_i$ parameters quantify the amount of spectral pinching. They encode information on the second-lowest spectral momentum, with
\begin{equation}
    \frac{\langle E_i^2\rangle}{\langle E_i\rangle^2} = \frac{2+\alpha_i}{1+\alpha_i}\,.
\end{equation}

We focus on the stages of the core-collapse supernova that follows the neutronization burst, specifically the accretion phase and the early cooling phase. During this time window, from several tens of milliseconds to a few seconds, the spectral characteristics of the neutrino flux are not expected to 
change abruptly, see e.g.~\cite{Hudepohl:2009tyy}.
As a first approximation, we take the pinching parameterization to remain valid during the analysis time window and consider constant (effective) values for the luminosities, pinching parameters, and average energies. 
Therefore, in what follows, we omit dots and discuss only time-integrated quantities.

\subsection{Phase-space distributions}

In order to evaluate the impact of collective effects, i.e.~of slow and fast flavour conversions, one must take into account the phase-space distribution functions $f_i$ of neutrinos and antineutrinos ($i = \nu_e,\,\bar\nu_e,\,\nu_\mu,\,\bar\nu_\mu,\,\nu_\tau,\,\bar\nu_\tau$). 
Prior to flavour conversions, we assume that non-electron-type neutrinos and antineutrinos share the same distribution, such that $f_{\nu_x} = f_{\bar\nu_x} $, where $f_{\nu_x} \equiv f_{\nu_\mu} = f_{\nu_\tau}$ and $f_{\bar\nu_x} \equiv  f_{\bar\nu_\mu} = f_{\bar\nu_\tau}$.
Moreover, we will assume that the phase-space distributions $f_i$ are separable, taking
\begin{equation} \label{eq:fsep}
f_i(v,E) = (2\pi)^2\, n_i \,\frac{f_i(v)}{\int f_i(v') \, dv'}\,\frac{f_i(E)}{\int f_i(E') \,E'^2\, dE'}\,,
\end{equation}
where $v= \cos \theta$ is the radial velocity (axisymmetric setup), 
while $n_i$ denotes the neutrino or antineutrino number density.
With the ansatz of~\cref{eq:fsep}, the normalization condition
\begin{equation}
\int_{-1}^{+1} \int_{0}^{+\infty} f_i(v,E)\, E^2 \,dE \,dv = (2\pi)^2\, n_i
\end{equation}
is automatically satisfied, independently of the normalizations of $f_i(v)$ and $f_i(E)$.
Note that, while this is sufficient for our phenomenological purposes,
a more realistic treatment of collective effects would rely on the output of SN simulations capable of fully reconstructing energy and angular information.

The total emitted energies corresponding to each neutrino species $\mathcal{L}_i^\text{tot}$ (hereafter referred to simply as `luminosity' for brevity), obtained as the time integral of the instantaneous luminosity $\mathcal{L}_i$, are related to the phase-space distributions via
\begin{equation} \label{eq:dLdEdef}
\frac{d\mathcal{L}_i^\text{tot}}{dE} = \frac{R^2\, E^3}{\pi} \int f_i(v,E) \, v\, dv \,,
\end{equation}
where $R$ is the neutrinosphere radius,%
\footnote{
For simplicity, we only consider phase-space distributions at the neutrinosphere.
Any energy dependency of the neutrinosphere (or decoupling) radius is understood to be encoded in the pinching of the spectrum~\cite{Janka:2017vlw}. 
We are ignoring differences among radii for the different (anti)neutrino species $i,j$ (see e.g.\mbox{\cite{Sawyer:2015dsa}}). Including them would modify~\mbox{\cref{eq:nratios}} by a factor $R_j^2 / R_i^2$.}
further implying
\begin{equation} \label{eq:fEs}
f_i(E) = \frac{(\alpha_i+1)^{\alpha_i+1}}{\langle E_i\rangle^3\, \Gamma(\alpha_i+1)}\left(\frac{E}{\langle E_i\rangle}\right)^{\alpha_i-2}\, e^{-(\alpha_i +1) \frac{E}{\langle E_i \rangle}}\,,
\end{equation}
where we have chosen the overall normalization constant such that $\int f_i(E)\,E^2\,dE = 1$.
Finally, taking~\cref{eq:dLdEdef} into account, we may parameterize the total luminosities as 
\begin{equation} \label{eq:dLdEparam}
\mathcal{L}^\text{tot}_i = 4\pi R^2\, n_i \langle E_i \rangle\langle v_i \rangle  \,\equiv\,
\epsilon_i \langle v_i \rangle\,,
\end{equation}
where $\langle v_i \rangle = \int_{-1}^{+1} f_i(v) \,v\,dv$ for an appropriately normalized $f_i(v)$,
so that the ratios of local number densities are given by
\begin{equation} \label{eq:nratios}
    \frac{n_i}{n_j} = \frac{\epsilon_i \langle E_j\rangle}{\epsilon_j \langle E_i\rangle}\,.
\end{equation}
The knowledge of these ratios is important to estimate flavour conversion effects, see next section.
Note that, in the case $\langle v_i \rangle = 1$, one has $\epsilon_i = \mathcal{L}^\text{tot}_i$ directly. Otherwise, for fixed $\epsilon_i$, the fluence and event rates are reduced by an overall factor of $\langle v_i \rangle < 1$.

Given the above, the spectral fluence at the Earth in the absence of flavour conversions (unoscillated fluence) reads 
\begin{equation} \label{eq:fluence}
    \frac{d\Phi_i}{dE} \,=\, \frac{1}{4\pi D^2}
     \frac{1}{E} \frac{d{\mathcal{L}}^\text{tot}_i}{dE}
    \,=\,
    \frac{1}{4\pi D^2} \frac{\epsilon_i \langle v_i \rangle}{\langle E_i \rangle^2} \frac{(\alpha_i +1)^{\alpha_i +1}}{\Gamma(\alpha_i +1)} \left(\frac{E}{\langle E_i\rangle}\right)^{\alpha_i}\, e^{-(\alpha_i +1) \frac{E}{\langle E_i \rangle}}
    \,,
\end{equation}
and is typically given in units of MeV$^{-1}$\,cm$^{-2}$. 
The luminosity of a single neutrino species, in a SN event is expected to be of order $\mathcal{L}_i^\text{tot} \sim \text{few} \times 10^{52}\,\text{erg}$, i.e.~a few tens of foe ($1\,\text{foe} \equiv 10^{51}\,\text{erg}$).
For $D \sim 10$ kpc, this corresponds to spectral fluences of the order $d\Phi_i/dE \sim 10^{10}\,\text{MeV}^{-1}\,\text{cm}^{-2}$.
We thus take~\cref{eq:fluence} as a parameterization of the flux prior to the action of flavour conversions, which we now discuss.

\subsection{Flavour conversions}
\label{sec:flacon}

The key quantities governing the onset of slow and fast flavour instabilities 
are the spectral differences
\begin{equation} \label{eq:g}
   g_{v,E} = \begin{cases}
       f_{\nu_e}(v,E) - 
       f_{\nu_x}(v,E) \,, & E>0\\[2mm]
       f_{\bar\nu_x}(v,|E|) - 
       f_{\bar\nu_e}(v,|E|) \,, & E<0
   \end{cases}\,,
\end{equation}
where the extension to negative $E$ allows to account for neutrinos and antineutrinos simultaneously.

\paragraph*{Slow instabilities.}
These instabilities develop in the presence of spectral crossings, that are in correspondence with zeros of the $v$-integrated differences
\begin{equation} \label{eq:gvE}
\begin{aligned}
g(1/E) \equiv
   \int dv\, g_{v,E} &\,\propto\, \begin{cases}
       n_{\nu_e} f_{\nu_e}(E) - 
       n_{\nu_x} f_{\nu_x}(E) \,, & E>0\\[2mm]
       n_{\nu_x} f_{\nu_x}(|E|) - 
       n_{\bar\nu_e} f_{\bar\nu_e}(|E|) \,, & E<0
   \end{cases}\\[2mm]
   &\,\propto\, \begin{cases}
       \alpha \,f_{\nu_e}(E) - 
      f_{\nu_x}(E) \,, & E>0\\[2mm]
       f_{\nu_x}(|E|) - 
       \bar\alpha \,f_{\bar\nu_e}(|E|) \,, & E<0
   \end{cases}\,,
\end{aligned}
\end{equation}
where we have defined 
$\alpha \equiv n_{\nu_e}/n_{\nu_x}$ and 
$\bar\alpha \equiv n_{\bar\nu_e}/n_{\nu_x}$,
and used $f_{\bar\nu_x}(v,E) = f_{\nu_x}(v,E)$. Taking into account the estimates of~\cref{eq:fEs,eq:nratios}, the above expression provides a handle on the position of spectral crossings $E_c$, and thus of spectral swaps and splits.%
\footnote{Note that, for $\langle v_i \rangle = 1$, the zero-crossings in $g$ are in correspondence with crossings of the $\nu_e$ and $\nu_x$ (or $\bar\nu_e$ and $\bar\nu_x$) fluxes, since $d\Phi_i /d E \propto n_i \langle v_i \rangle E^2 f_i(E)$.}
The stability of these crossings depends on the slope of $g$ at $1/E_c$ and on the ordering of the neutrino mass spectrum, which can be normal (NO) or inverted (IO). A crossing with positive (negative) slope is unstable in the IO (NO) case, and will produce a spectral swap across a range of energies around $E_c$~\cite{Dasgupta:2009mg} (see also~\cite{Dasgupta:2010cd}). 
Spectral splits correspond to the limits $E_c \pm \Delta E$ of this energy window, which we assume to be symmetric for simplicity.
Moreover, we take these splits to be sharp as a rough approximation, while in reality the swap factor relating pre- and post-conversion fluxes is expected to be smooth, becoming more complex in the presence of multi-angle effects.%
\footnote{Recent studies indicate that time-dependent fluctuations~\cite{Dasgupta:2015iia,Abbar:2015fwa} may play an important role, potentially washing out spectral swaps. Moreover, spectral crossings might not be sufficient for the latter to develop~\cite{Fiorillo:2025kko}. Instead, the evolution of slow instabilities could lead to a turbulent bath of flavour waves and to an approximate equipartition among antineutrino species~\cite{Fiorillo:2024pns,Fiorillo:2025ank}. In this work, we consider the extreme scenario in which spectral swaps are fully realized, in order to investigate their maximal phenomenological impact.}

In line with the results of SN simulations, we take $\alpha_{\nu_x} < \alpha_{\nu_e,\,\bar\nu_e}$ and $E_{\nu_x} > E_{\nu_e,\,\bar\nu_e}$ (see also the next section). It follows that the slope of $g$ at the origin is negative, leading to a zero-crossing at $|E_c| = \infty$. This crossing is unstable in the NO case, producing a swap window shared by the neutrino and antineutrino spectra. Therefore, for NO, a swap occurs in both spectra for energies above a certain split value, $|E| > E_\text{sp}^\text{NO}$.
In practice, following e.g.~\cite{Abbar:2024nhz},
we fix the window half-width to be $\Delta E = 7.5$~MeV, while the aforementioned split value is set at $E_\text{sp}^\text{NO} = 20$~MeV. Overlapping swap windows are taken to merge, since multiple close crossings can act as a single one~\cite{Dasgupta:2009mg}.

Given the parameterization in~\cref{eq:fEs}, the positions of the (finite) crossings $E_c$ can be determined analytically by solving 
\begin{equation}
    \alpha \,f_{\nu_e}(E_c) = f_{\nu_x}(E_c)\qquad \text{and} \qquad
    \bar\alpha \,f_{\bar\nu_e}(|E_c|) = f_{\nu_x}(|E_c|)\,.
\end{equation}
When pinching parameters differ, $\alpha_{\nu_x} \neq \alpha_{\nu_e,\,\bar\nu_e}$,
these equalities can be cast in the form $y\, e^y=x$, with $y = w\,|E_c|$ and 
\begin{equation}
x = w
\left(
\frac{\alpha_{\nu_e}+1}{\langle E_{\nu_e}\rangle}
\right)^{\frac{1+\alpha_{\nu_e}}{\alpha_{\nu_x}-\alpha_{\nu_e}}}
\left(
\frac{\alpha_{\nu_x}+1}{\langle E_{\nu_x}\rangle}
\right)^{-\frac{1+\alpha_{\nu_x}}{\alpha_{\nu_x}-\alpha_{\nu_e}}}
\left[\alpha\,
\frac{\Gamma(\alpha_{\nu_x}+1)}{\Gamma(\alpha_{\nu_e}+1)}
\right]^{\frac{1}{\alpha_{\nu_x}-\alpha_{\nu_e}}}
\,,
\end{equation}
for neutrino ($E_c>0$) crossings,
where
\begin{equation}
w = \frac{1}{\alpha_{\nu_x}-\alpha_{\nu_e}}\left(\frac{\alpha_{\nu_e}+1}{\langle E_{\nu_e}\rangle}-\frac{\alpha_{\nu_x}+1}{\langle E_{\nu_x}\rangle}\right)\,.
\end{equation}
Analogous expressions hold for antineutrino crossings ($E_c<0$), with the replacements
$\alpha_{\nu_e} \to \alpha_{\bar\nu_e}$,
$\langle E_{\nu_e}\rangle \to \langle E_{\bar\nu_e}\rangle$
and $\alpha \to \bar \alpha$.
A crossing exists provided $x \geq -1/e$, and two crossings are present only if additionally $x<0$. The crossing energies are then given by $|E_c|= y/ w$, with $y = W_0(x)$, or
$y = W_0(x)$ and $y = W_{-1}(x)$ in the two-crossing case, where $W$ is the product logarithm (or Lambert) function.
In the limit of equal pinching parameters, say  $\alpha_{\nu_e} = \alpha_{\nu_x}$ while keeping $\langle E_{\nu_e} \rangle \neq \langle E_{\nu_x} \rangle$, one directly finds
\begin{equation}
|E_c| = \frac{\langle E_{\nu_e} \rangle\langle E_{\nu_x} \rangle}{\langle E_{\nu_x} \rangle-\langle E_{\nu_e} \rangle}\left(\frac{\log \alpha}{\alpha_{\nu_e}+1}+\log\frac{\langle E_{\nu_x} \rangle}{\langle E_{\nu_e} \rangle}\right)
\end{equation}
instead. The antineutrino case is recovered via the aforementioned replacements.
In any case, the pinched neutrino spectrum will admit at most two spectral crossings at finite energy and only one of these will be unstable (the same holds for antineutrinos).

\paragraph*{Fast instabilities.}
These instabilities lead to energy-independent FFCs and develop in the presence of angular crossings. 
These crossings are in correspondence with the zeros of the difference between electron-neutrino lepton number (ELN) and non-electron-neutrino lepton number (XLN) angular distributions, proportional to
\begin{equation}
\begin{aligned}
 G(v) &\equiv \int_{-\infty}^{+\infty} g_{v,E} \,E^2 \,dE \\
 &= \int_0^{+\infty} \Big[
 \big(f_{\nu_e}(v,E)-f_{\bar\nu_e}(v,E)\big)
 -
 \big(f_{\nu_x}(v,E)-f_{\bar\nu_x}(v,E)\big)
 \Big] E^2\,dE 
 \,,
\end{aligned}
\end{equation}
obtained from integrating~\cref{eq:g} over (anti)neutrino energies. In the simpler case where initially $f_{\bar\nu_x}(v,E) = f_{\nu_x}(v,E)$, one finds
\begin{equation} \label{eq:Gsimple}
      G(v) \,\propto\,   f_{\nu_e}(v) - \beta\, f_{\bar\nu_e}(v)
  \,,
\end{equation}
with $\beta \equiv n_{\bar\nu_e}/n_{\nu_e} = \bar\alpha/\alpha$ and 
the $f_i(v)$ henceforth normalized as
$\int_{-1}^{+1} f_i(v)\,dv = 1$.
In the more general case, with $f_{\bar\nu_x} \neq f_{\nu_x}$, one has instead
\begin{equation} \label{eq:Ggen}
\begin{aligned}
      G(v) &\,\propto\,  n_{\nu_e}\, f_{\nu_e}(v) - n_{\nu_x}\, f_{\nu_x}(v) + n_{\bar\nu_x}\, f_{\bar\nu_x}(v) - n_{\bar\nu_e}\, f_{\bar\nu_e}(v)
      \\[2mm]
      &\,\propto\, f_{\nu_e}(v) - \gamma\, f_{\nu_x}(v) + \bar\gamma\, f_{\bar\nu_x}(v) - \beta\, f_{\bar\nu_e}(v)
  \,,
\end{aligned}
\end{equation}
with $\gamma \equiv  n_{\nu_x}/n_{\nu_e} = 1/\alpha$ and $\bar\gamma \equiv  n_{\bar\nu_x}/n_{\nu_e}$.

The ensuing flavour conversion (or depolarization~\cite{Bhattacharyya:2020dhu,Bhattacharyya:2020jpj,Bhattacharyya:2022eed})
proceeds in a $v$-depen\discretionary{-}{}{}dent fashion.
Following the prescription detailed in Ref.~\cite{Zaizen:2022cik} (see also~\cite{Xiong:2023vcm}), the extent to which electron-flavour and heavier-flavour neutrinos are equilibrated 
depends on the survival probability $P_{ee}(v)$, approximated by the step function 
\begin{equation} \label{eq:Pee}
   P_{ee}(v) = \begin{cases}
       p_\text{eq} \,, & G(v)<0\\[2mm]
       1-(1-p_\text{eq})\, \dfrac{A}{B} \,, & G(v)>0
   \end{cases}\,,
\end{equation}
in the case of a positive $\text{ELN}-\text{XLN} \propto \int G(v)\, dv$, i.e.~$B-A > 0$, with
\begin{equation}
    A \,\equiv\, \left| \int_{G(v)<0} G(v)\, dv\right|\,,\qquad
    B \,\equiv\, \left| \int_{G(v)>0} G(v)\, dv\right|\,.
\end{equation}
If instead $B-A < 0$, one finds
\begin{equation} \label{eq:Peeneg}
   P_{ee}(v) = \begin{cases}
       1-(1-p_\text{eq})\, \dfrac{B}{A} \,, & G(v)<0\\[2mm]
       p_\text{eq} \,, & G(v)>0
   \end{cases}\,.
\end{equation}
Note that this simple result is based on numerical simulations of homogeneous local boxes. For a more complete theoretical treatment based on quasi-linear theory, see~\cite{Fiorillo:2024qbl}.
In the scheme we are considering, crossings are dynamically eliminated via i) full equilibration in the angular region with smaller (absolute) integral, accompanied by ii) a reduction of the absolute integral in the other region, dictated by ELN and XLN conservation.
For the full equilibration region, one finds $p_\text{eq} = 1/3$ in the three-flavour case.%
\footnote{See~\cite{Liu:2025tnf} for a generalization considering possible differences among non-electron (anti)neutrinos.}
Note that in deriving~\cref{eq:Pee} a single initial ELN$-$XLN crossing was assumed.
In our implementation we tentatively take these results to hold also in the presence of multiple crossings, while awaiting validation from quantum kinetic simulations.

As a result of FFCs, the phase-space distributions $f_i(v,E)$ are modified as
\begin{equation} \label{eq:FFCfs}
\begin{aligned}
    f_\nuebp^\text{FFC} & \,=\, P_{ee}\, f_\nuebp^0 + (1-P_{ee})\, f_\nuxbp^0\,,
    \\[2mm]
    f_\nuxbp^\text{FFC} & \,=\, \frac{1-P_{ee}}{2}\, f_\nuebp^0 + \frac{1+P_{ee}}{2}\, f_\nuxbp^0
    \,,
\end{aligned}
\end{equation}
where the `0' superscript identifies the distributions before the development of FFCs,
leading to the updated fluences
\begin{equation} \label{eq:FFCfluences}
\begin{aligned}
    \frac{d\Phi_\nuebp^\text{FFC}}{dE} & \,=\, 
    \frac{    \int  P_{ee}(v)\, f_\nuebp(v)\, v\,dv }{\langle v_\nuebp\rangle}\,
    \frac{d\Phi_\nuebp^0}{dE}
    +
    \frac{    \int  \big(1-P_{ee}(v)\big) f_\nuxbp(v)\, v\,dv }{\langle v_\nuxbp\rangle}\,
    \frac{d\Phi_\nuxbp^0}{dE}
    \,,
    \\[2mm]
    \frac{d\Phi_\nuxbp^\text{FFC}}{dE} & \,=\, 
    \frac{    \int \big(1-P_{ee}(v)\big) f_\nuebp(v)\, v\,dv }{2\,\langle v_\nuebp\rangle}\,
    \frac{d\Phi_\nuebp^0}{dE}
    +
    \frac{    \int  \big(1+P_{ee}(v)\big) f_\nuxbp(v)\, v\,dv }{2\,\langle v_\nuxbp\rangle}\,
    \frac{d\Phi_\nuxbp^0}{dE}
    \,,
\end{aligned}
\end{equation}
where we have used the fact that survival probabilities are energy-independent in the FFC regime and thus coincide for neutrinos and antineutrinos.

In what follows, we model the $v$ dependencies as Gaussian distributions centred in the forward direction,
\begin{equation} \label{eq:Gauss}
    f_i(v) = \underbrace{\left[\sqrt{\frac{\pi}{2}}\,\sigma_i\erf\left(\frac{\sqrt{2}}{\sigma_i}\right)\right]^{-1}}_{ \equiv\,
    \pazocal{N}_i} \exp\left[-\frac{(v-1)^2}{2\sigma_i^2}\right]
    \,,
\end{equation}
where the $\pazocal{N}_i$ pre-factors guarantee an appropriate normalization (note that other choices for $f_i(v)$ are possible, e.g.~a maximum entropy distribution~\cite{Richers:2022dqa,Xiong:2023vcm}).
Then, in the simpler case with $f_{\bar\nu_x}(v,E) = f_{\nu_x}(v,E)$,
the difference $G(v)$ is given by~\cref{eq:Gsimple} and presents at most one zero-crossing $v_c \in [-1,1]$, given by
\begin{equation}
v_c
=
 1-\sqrt{2}\,\sigma_{\nu_e}\sigma_{\bar\nu_e}\sqrt{\frac{
\log (\beta \,\pazocal{N}_{\bar\nu_e}/\pazocal{N}_{\nu_e}) 
}{\sigma_{\nu_e}^2-\sigma_{\bar\nu_e}^2}}
\simeq 1-\sqrt{2}\,\sigma_{\nu_e}\sigma_{\bar\nu_e}\sqrt{\frac{
\log \beta + \log (\sigma_{\nu_e}/\sigma_{\bar\nu_e}) 
}{\sigma_{\nu_e}^2-\sigma_{\bar\nu_e}^2}}
\,,
\end{equation}
provided $v_c \in \mathbb{R}$, where the approximation holds for $\sigma_{\nu_e,\,\bar\nu_e}< 1$.
Instead, in the more general case with $f_{\bar\nu_x}\neq f_{\nu_x}$ and given the choice in~\cref{eq:Gauss},
one can have up to 3 crossings of~\cref{eq:Ggen} in the interval $[-1,1]$, 
which must instead be determined numerically.

\vskip 2mm
One may envision conversion scenarios where only slow (fast) flavour conversions are active, while fast (slow) ones remain disabled. For instance, it suffices to take $f_i(v) \propto \delta(v-1)$ to consistently turn off FFC effects in this setup, while allowing the triggering of slow instabilities due to spectral crossings.
Notice, however, that slow (fast) flavour conversions alter the angular (spectral) distributions, so care must be taken when constructing a phenomenological pipeline that incorporates both types of conversion effects.%
\footnote{One should keep in mind that the sequential application of slow and fast flavour conversions is still a phenomenologically driven simplification of a potentially more complex interplay of effects.}
There are recent indications~\cite{Fiorillo:2025gkw} that slow instabilities in the SN can sufficiently develop and shape the flavour pattern before fast ones arise. Therefore, in the joint scenario where both fast and slow instabilities are active, we consider that FFCs take place after spectral swaps. 
To this end, we appropriately modify~\cref{eq:FFCfs,eq:FFCfluences}, applying sequentially the effects of slow and fast flavour conversions, taking into account that distributions are no longer separable and expressing our results as functions of the initial parameters.

\paragraph*{MSW conversions and vacuum propagation.}
Following the action of collective effects, (anti)neutrinos experience a decreasing matter density profile as they escape the SN environment and travel through vacuum toward the Earth. At high matter density, flavour states approximately coincide with matter eigenstates and, provided adiabaticity is maintained (i.e.~in the absence of shocks~\cite{Fogli:2003dw,Tomas:2004gr} or turbulence~\cite{Kneller:2010sc,Borriello:2013tha}), they evolve into mass eigenstates in vacuum. Therefore, as they arrive at the detector, their flavour content is determined by the vacuum mixing angles. One then has~\cite{Dighe:1999bi}, denoting with the superscript `0' the distributions before the MSW effects take place:%
\footnote{In the presence of collective effects, which occur substantially before the standard MSW effects, ${d\Phi_{\nu_e}^0}/{dE}$ are the distributions after the development of slow and fast flavour conversions and therefore they do not coincide with those used in the right-hand side of~\cref{eq:FFCfluences}.}
\begin{equation} \label{eq:MSWfluences}
\begin{aligned}
    \frac{d\Phi_{\nu_e}^\text{MSW}}{dE} & \,=\, 
    p\,\frac{d\Phi_{\nu_e}^0}{dE}
    +
    (1-p)\,
    \frac{d\Phi_{\nu_x}^0}{dE}
    \,,
    \\[2mm]
    \frac{d\Phi_{\nu_x}^\text{MSW}}{dE} & \,=\, 
    \frac{1-p}{2}\,    \frac{d\Phi_{\nu_e}^0}{dE}
    +
    \frac{1+p}{2}\,    \frac{d\Phi_{\nu_x}^0}{dE}
    \,,
\end{aligned}\quad
\begin{aligned}
    \frac{d\Phi_{\bar\nu_e}^\text{MSW}}{dE} &\,=\, 
    \bar{p}\,\frac{d\Phi_{\bar\nu_e}^0}{dE}
    +
    (1-\bar{p})\,
    \frac{d\Phi_{\bar\nu_x}^0}{dE}
    \,,
    \\[2mm]
    \frac{d\Phi_{\bar\nu_x}^\text{MSW}}{dE} &\,=\, 
    \frac{1-\bar{p}}{2}\,    \frac{d\Phi_{\bar\nu_e}^0}{dE}
    +
    \frac{1+\bar{p}}{2}\,    \frac{d\Phi_{\bar\nu_x}^0}{dE}
    \,,
\end{aligned}
\end{equation}
where, crucially, $p$ and $\bar{p}$ depend on the neutrino mass ordering, with
\begin{equation} \label{eq:MSWprobs}
   p = \begin{cases}
       \sin^2 \theta_{13} \,, & \text{NO}\\[2mm]
       \sin^2 \theta_{12} \cos^2 \theta_{13} \,, & \text{IO}
   \end{cases}\,,
   \,\qquad\,
   \bar{p} = \begin{cases}
       \cos^2 \theta_{12} \cos^2 \theta_{13} \,, & \text{NO}\\[2mm]
       \sin^2 \theta_{13} \,, & \text{IO}
   \end{cases}\,.
\end{equation}
It is worth mentioning that applying~\cref{eq:MSWfluences,eq:MSWprobs} directly to the post-collective flavour fluxes is not strictly correct since the high-density matter eigenstates ${\nu_i^m}$ may differ non-trivially from flavour eigenstates~\cite{Capanema:2024hdm}. However, since we adopt an effective two-flavour description throughout, this subtlety does not affect our results.
Finally, although Earth-matter effects are not included in our analysis, they might give rise to observable features in the high-energy (low-statistics) region of the spectrum, see e.g.~\cite{Hajjar:2023knk}.

\subsection{Benchmark scenarios and priors}
\label{sec:benchmarks}

In order to assess the impact of different flavour conversion scenarios on the (anti)neutrino signal, we consider two explicit benchmarks for the initial fluxes, 
\A and \B, defined in~\cref{tab:benchmarks}. These are described by parameters pertaining to luminosity, $\mathcal{L}^\text{tot}_i = \langle v_i \rangle\,\epsilon_i$, spectral pinching ($\alpha_i$), average energy ($\langle E_i\rangle$) and angular spread ($\sigma_i$), characterizing the distribution of each species (recall that we take  $f_{\nu_x} = f_{\bar\nu_x} $ prior to flavour conversions).
As can be seen from the table, benchmarks differ only in the choice of pinching parameters, with $\alpha_i$ = 2.5 across all species in benchmark \A.
For definiteness, we also fix the Earth-SN distance to be $D=10$~kpc, keeping in mind that fluxes and event rates scale as $D^{-2}$.

%%%%%%%%%%%%%%%%%%%%%%%%%%%%%%%%%%%%%%
\begin{table}[t!]
  \centering
  \begin{tabular}{lcccccc}
    \toprule
    Benchmark & Species & $\epsilon_i$ ($10^{52}$ erg) & $\quad\alpha_i\quad$ & $\langle E_i \rangle$ (MeV) & $\quad\sigma_i\quad$ & $\quad\langle v_i\rangle \quad$\\
    \midrule
    & $\nu_e$ & 2.0 & 2.5 & 9.5 & 0.65 & 0.48 \\
    \A & $\bar\nu_e$  & 2.0 & 2.5 & 12.0 & 0.50 & 0.60 \\
    & $\nu_x$ & 2.0 & 2.5 & 15.6 & 0.40 & 0.68 \\
    \midrule
    & $\nu_e$ & 2.0 & 3.5 & 9.5 & 0.65 & 0.48 \\
    \B & $\bar\nu_e$  & 2.0 & 4.2 & 12.0 & 0.50 & 0.60  \\
    & $\nu_x$ & 2.0 & 3.0 & 15.6 & 0.40 & 0.68  \\
    \bottomrule
  \end{tabular}\\[2mm]
  Priors:\quad $\langle E_{\nu_e} \rangle \leq 
  \langle E_{\bar\nu_e} \rangle \leq \langle E_{\nu_x} \rangle$,
  \quad
   $\alpha_{\nu_x} \leq 
  \alpha_{\nu_e} \leq \alpha_{\bar\nu_e}$
  \caption{Pre-conversion spectral and angular parameters defining the considered benchmarks ($\nu_x$ and $\bar\nu_x$ distributions coincide).}
  \label{tab:benchmarks}
\end{table}
%%%%%%%%%%%%%%%%%%%%%%%%%%%%%%%%%%%%%%

The relevant oscillation parameters, i.e.~the leptonic mixing angles in vacuum, are fixed to $\sin^2\theta_{12} = 0.304$ and $\sin^2\theta_{13} = 0.0222$, within the $1\sigma$ ranges identified in recent global fits~\cite{Capozzi:2025wyn,Capozzi:2025ovi}.
In practice, as priors within our analysis, parameters are bound within the ranges
\begin{equation}
\epsilon_i \in [0,20]\times10^{52}\,\text{erg}\,,\quad
\alpha_i \in [2,10]\,,\quad
\langle E_i \rangle \in [5,25]\,\text{MeV}
\,.
\end{equation}
Inspired by the Garching model~\cite{Hudepohl:2009tyy},
we further impose the inequality priors on average energies and pinching parameters summarized below~\cref{tab:benchmarks}. 
While we consider fixed $\sigma_i$ in our analysis for simplicity, the inequality $\sigma_{\nu_x} \leq  \sigma_{\bar\nu_e} \leq \sigma_{\nu_e}$ among angular spreads is respected, in line with e.g.~\cite{Tamborra:2017ubu,Shalgar:2019kzy}.
Finally, the prior $|\log_{10}\epsilon_i/\epsilon_j| < 1$ is imposed on the $\epsilon_i$ luminosity parameters,
requiring them to be within one order of magnitude of each other.

%%%%%%%%%%%%%%%%%%%%%%%%%%%%%%%%%%%%%%%%%%%%%%
\section{Supernova neutrinos at DUNE}
\label{sec:DUNE}
%%%%%%%%%%%%%%%%%%%%%%%%%%%%%%%%%%%%%%%%%%%%%%

The Deep Underground Neutrino Experiment (DUNE)~\cite{DUNE:2015lol,DUNE:2020jqi,DUNE:2020ypp} is among the most promising next-generation neutrino experiments.%
\footnote{Other notable upcoming neutrino oscillation experiments which might provide valuable data on supernova neutrinos include Hyper-K~\cite{Hyper-Kamiokande:2018ofw} and JUNO~\cite{JUNO:2023dnp}.}
Its primary goal is to study accelerator-produced neutrino oscillations using an intense, wide-band muon-neutrino beam with a peak energy around 2.5 GeV. The accelerator complex and the Near Detector (ND) system~\cite{DUNE:2021tad} will be located at Fermilab, while the Far Detector (FD) will consist of four 10 kt fiducial-mass liquid argon time projection chamber (LArTPC) modules installed at the Sanford Underground Research Facility (SURF) in South Dakota, approximately 1300 km from the neutrino source.
Thanks to the large fiducial mass of the FD and the excellent imaging and calorimetric capabilities of the LArTPC technology, DUNE will not only perform precision measurements of accelerator neutrino oscillations but will also enable a broad non-beam neutrino physics programme~\cite{Capozzi:2018dat,DUNE:2020ypp,DUNE:2020zfm,Kelly:2021jfs}. In particular, its sensitivity to neutrino energies down to about 5 MeV~\cite{DUNE:2020zfm} makes it a powerful observatory for supernova neutrinos, providing a unique opportunity to study neutrino emission from a core-collapse event within our Galaxy.

As widely discussed in the previous sections, the supernova neutrino burst consists of three main components: electron neutrinos ($\nu_e$), electron antineutrinos ($\bar{\nu}_e$), and the remaining muon and tau (anti)neutrino species ($\nu_x$, $\bar{\nu}_x$). The $\nu_e$ and $\bar{\nu}_e$ components can undergo charged-current (CC) interactions, while $\nu_x$ and $\bar{\nu}_x$, being typically below the energy threshold required for muon or tau production, interact only through neutral-current (NC) processes.
Consequently, the main detection channels for supernova neutrinos in DUNE are~\cite{DUNE:2020zfm,Saez:2023snv}:
\begin{itemize}
    \item $\nu_e$ CC interactions with argon nuclei,
    \begin{equation}
        \nu_e \, + \, ^{40}\mathrm{Ar} \to \, e^- \, + \, ^{40} \mathrm{K}^*\,,
    \end{equation}
    where the observable signature arises from the emitted electron and the subsequent de-excitation of the daughter nucleus.
    \item $\bar\nu_e$ CC interactions with argon nuclei,
    \begin{equation}
        \bar\nu_e \, + \, ^{40}\mathrm{Ar} \to \, e^+ \, + \, ^{40} \mathrm{Cl}^*\,,
    \end{equation}
   in which the visible energy originates from the positron and the de-excitation photons from the final-state nucleus.  
    \item NC interactions with argon nuclei,
    \begin{equation}
        \nu \, + \, ^{40}\mathrm{Ar} \to \, \nu \, + \, ^{40} \mathrm{Ar}^*\,,
    \end{equation}
    where $\nu$ denotes any neutrino or antineutrino flavour. In this case, the only detectable signal is due to the de-excitation of the recoiling argon nucleus.  
    \item Neutrino-electron elastic scattering (ES),
    \begin{equation}
        \nu \, + \, e^- \to \, \nu \, + \, e^-\,,
    \end{equation}
    which proceeds via Z-boson exchange (NC) for all flavours, and additionally via W-boson exchange (CC) for $\nu_e$. The observable in this channel is the energy of the recoil electron.
\end{itemize}
In our analysis, we considered as signal all events arising from the four detection channels described above, neglecting background contributions for simplicity. Such backgrounds are expected to be negligible and mainly originate from misidentified events, cosmogenic and radiological sources~\cite{DUNE:2020zfm}. Moreover, we do not consider pile-up effects due to the large amount of neutrinos arriving at the detector in a short time, which would need a dedicated analysis.
To compute the expected event rates, 
we employed the General Long Baseline Experiment Simulator (GLoBES)~\cite{Huber:2004ka,Huber:2007ji}
with a custom probability engine and appropriately modified flux and AEDL files.
We have used the smearing matrices
provided within the SNOwGLoBES package~\cite{SnowGLOBES} in the 40 kt LArTPC case,
which model the DUNE detector response for each interaction channel.
As a first-order approximation, we adopted for all channels the detection efficiency function reported in~\cite{DUNE:2020zfm}, originally derived for $\nu_e$ CC interactions. Furthermore, we considered only events with true neutrino energies above 5 MeV as detectable. For the sake of simplicity, we have neglected systematic uncertainties, which would require a detailed separate discussion, see e.g.~\cite{DUNE:2020zfm,DUNE:2023rtr}.

\begin{figure}[t]\hspace{-1.6cm}
    \includegraphics[height=7cm]{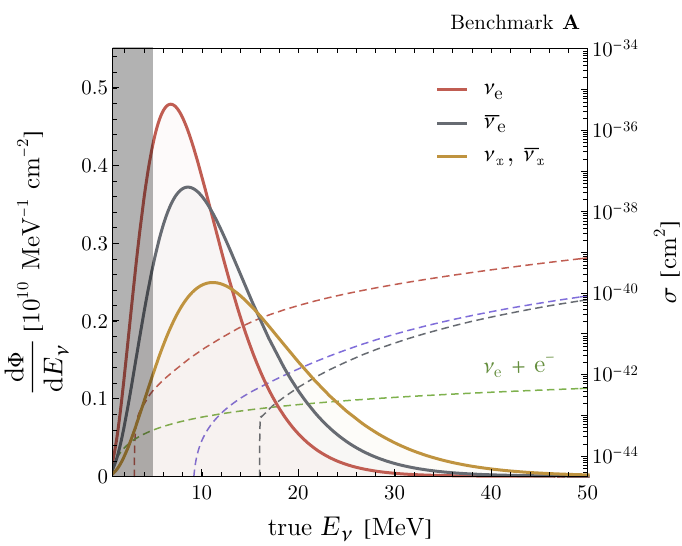}\hspace{4mm}
    \includegraphics[height=7cm]{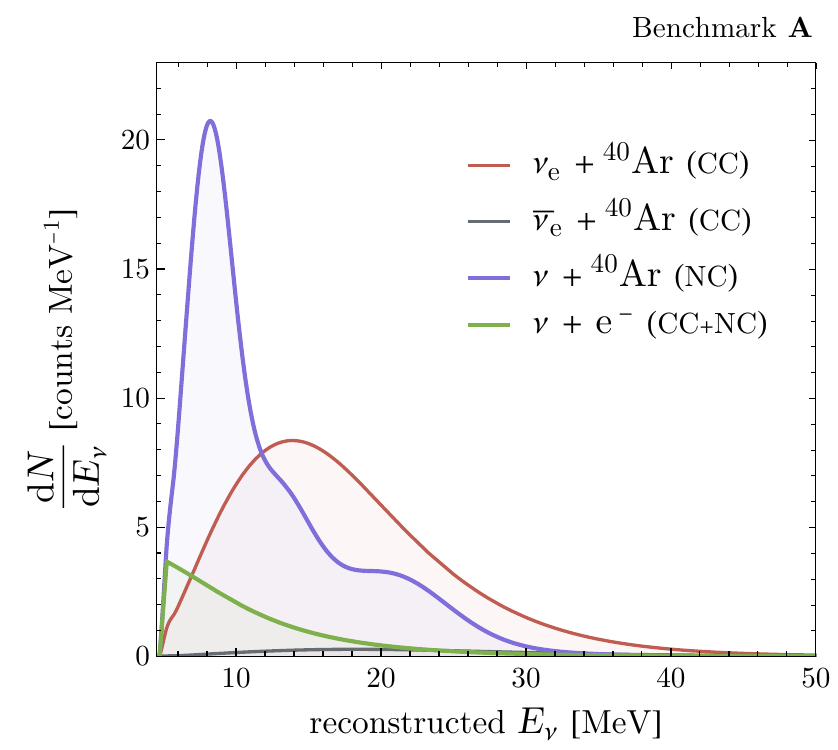}
\vskip 2mm
    \caption{Spectral fluences and cross sections for the different (anti)neutrino flavours, as functions of their true energies (left), and expected event rates in DUNE for each detection channel, as functions of reconstructed energy (right), for benchmark {\A} and in the absence of flavour conversions.}
    \label{fig:events}
\end{figure}

In~\cref{fig:events} (left panel), we show the fluences corresponding to
the parameters of benchmark \A,%
\footnote{We do not show here the corresponding plots for benchmark \B, 
as they share very similar features with those in~\cref{fig:events}.}
defined in~\cref{sec:benchmarks}.
The red, grey, and dark yellow lines represent the $\nu_e$, $\bar{\nu}_e$, and $\nuxbp$ fluences, respectively. The same plot also includes, as dashed lines, the cross sections adopted in our simulations for the four detection channels taken from the SNOwGLoBES package~\cite{SnowGLOBES}:%
\footnote{See~\cite{DUNE:2023rtr} for a detailed discussion on the impact of different cross-section models on expected supernova neutrino spectra and on parameter estimation at DUNE.}
red for $\nu_e$ CC, grey for $\bar{\nu}_e$ CC, blue for NC, and green for ES interactions. Some considerations are in order. The largest cross section corresponds to the $\nu_e$ CC interaction, while the NC and $\bar{\nu}_e$ CC cross sections are comparable, although the latter features a significantly higher energy threshold. The ES channel exhibits the smallest cross section, despite being the only one without a minimum neutrino energy requirement. Note that the ES cross section shown in the figure corresponds exclusively to the electron-neutrino flavour, which is higher than that for other flavours due to the additional CC contribution.

The right panel of~\cref{fig:events} displays the event spectra obtained with the GLoBES software for all channels. Owing to its large cross section and low threshold, the $\nu_e$ CC spectrum (red line) closely follows the shape of the incident $\nu_e$ fluence. In contrast, the $\bar{\nu}_e$ CC spectrum (grey line) is strongly suppressed, as only a small fraction of $\bar{\nu}_e$ are above the interaction threshold. The NC spectrum (blue line), which includes contributions from all neutrino flavours, peaks at relatively low reconstructed energies, around 10 MeV. This behaviour can be explained by the fact that, in these interactions, the visible energy originates only from nuclear de-excitation, leading to a systematic underestimation of the reconstructed neutrino energy. Similar considerations apply to the ES channel (green line), which contributes only marginally to the total event rate in DUNE, due both to its small cross section and the difficulty in reconstructing the true neutrino energy from the electron recoil signal.

%%%%%%%%%%%%%%%%%%%%%%%%%%%%%%%%%%%%%%%%%%%%%%
\section{Results}
\label{sec:results}
%%%%%%%%%%%%%%%%%%%%%%%%%%%%%%%%%%%%%%%%%%%%%%
\begin{figure}[p]
\centering
    \includegraphics[width=15cm]{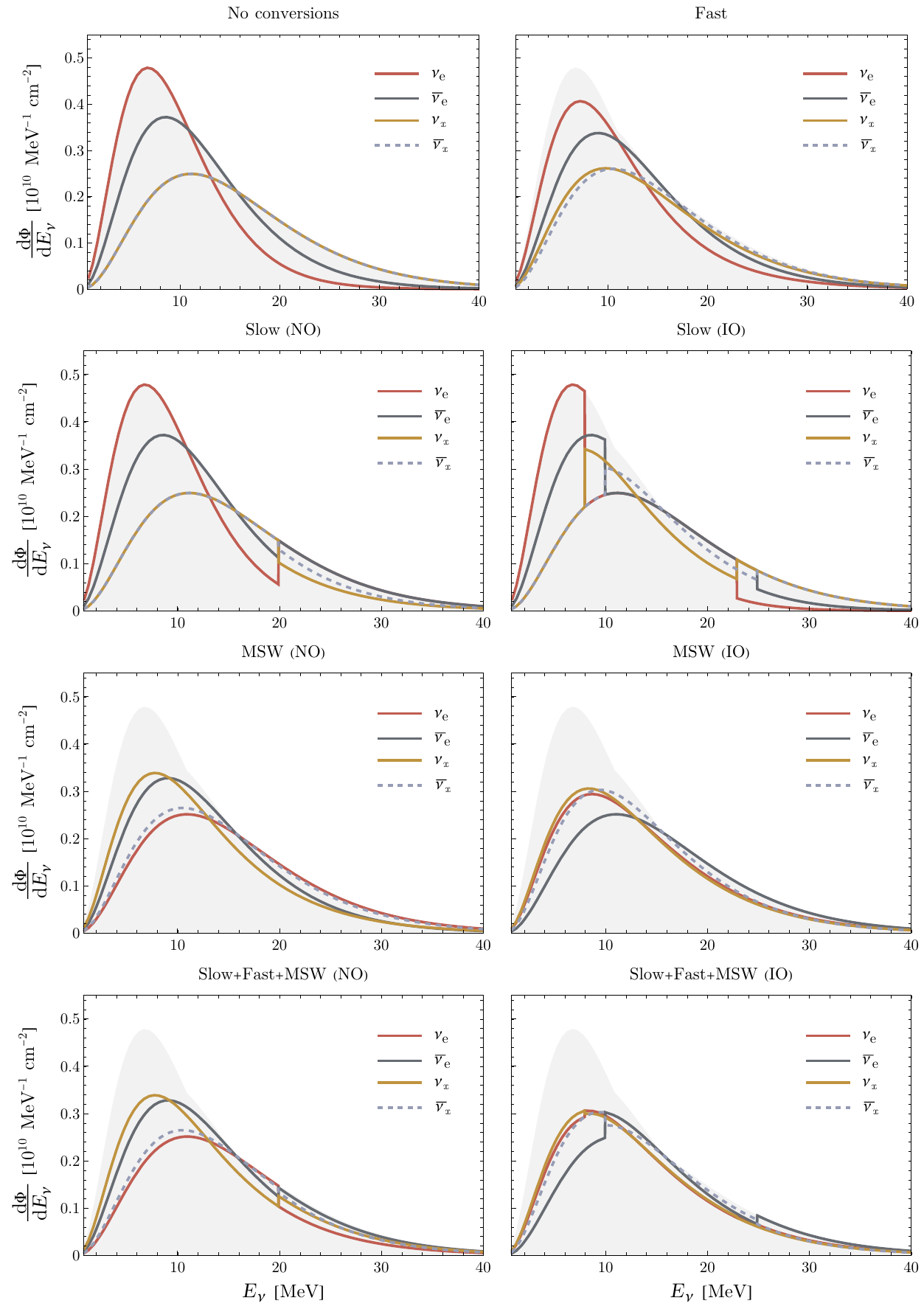}
    \caption{
    Spectral fluences for the different (anti)neutrino flavours, as functions of their true energies, for benchmark {\A} and the different, indicated flavour conversion scenarios.
    Grey shading corresponds to the reference unoscillated case.
    }
    \label{fig:AfluencesNOIO}\vspace{-60pt}
\end{figure}

In~\cref{fig:AfluencesNOIO}, we showcase the effects of the different flavour conversion scenarios on the original spectral fluences, for benchmark \A. We consider separately each mechanism, i.e.~fast, slow, and MSW flavour conversions, as well as their cumulative effect.
Note that, unlike MSW effects or the spectral swaps induced by slow instabilities, FFCs by themselves are independent of the neutrino mass ordering.
In the considered benchmark, the $\nu_e$ fluence is predominantly peaked at low energies, while the other fluences exhibit broader distributions, with $\nu_x$ remaining significant up to 40 MeV.

Fast conversions alone tend to equalize the fluences across different neutrino species, diminishing the distinctions between them. However, the differing fluences of $\nu_e$ and $\bar\nu_e$, which mix with $\nu_x$ and $\bar\nu_x$ respectively, allow for the separation of neutrino and antineutrino fluences for non-electron species. These differences may nonetheless be subtle and challenging to observe experimentally.

The standard MSW effects are qualitatively similar to fast conversions but more pronounced, owing to the small value of the $\theta_{13}$ mixing angle, which suppresses the $\nu_e$ fluence in favour of $\nu_x$ in the NO case, while in the IO case it is the $\bar\nu_e$ fluence that experiences a characteristic reduction in favour of $\bar\nu_x$. This highlights how supernova neutrinos could play a role in determining the neutrino mass ordering, independently from accelerator and reactor neutrino oscillation experiments.

By contrast, slow conversions can have more dramatic consequences. In our benchmark case, spectral swaps induce transitions around specific energies. For NO, both $\nu_e$ and $\bar\nu_e$ fluences increase above $E_\text{sp}^\text{NO} = 20$~MeV, while those of other species decrease. Given that certain detection channels in DUNE favour $\nu_e$ observation, these differences could be amplified in experimental signatures. For IO, two spectral crossings emerge, creating even more striking features. Notably, the first swap at approximately 8 MeV causes a substantial depletion of the $\nu_e$ fluence, since in that energy region the $\nu_e$ fluence exceeds the $\nu_x$ fluence by a factor of two. Such spectral features, if they occur, should be readily observable.

When all three effects are combined, partial compensation occurs. The dramatic fluence modifications from slow conversions are softened by fast instabilities, which drive the fluences toward equality. The subsequent MSW effect may then amplify certain differences, partially enabling the observation of specific fluence variations.

\vskip 2mm
The main conclusion from this preliminary fluence analysis is that neglecting to incorporate collective oscillation effects in the analysis of supernova neutrino spectra risks substantial misinterpretation of the original neutrino flux.
Even within this simplified, phenomenological framework, multiple competing effects clearly emerge during neutrino propagation toward Earth, and inadequate modelling could severely limit our ability to extract the rich information these particles carry. While the presence of resolved spectral swaps might serve as a clear signature of slow conversions, their absence does not, per se, pinpoint the underlying neutrino flavour conversion mechanism(s).

\vskip 2mm
In what follows,
we focus on three different flavour conversion scenarios:
\begin{itemize}
    \item MSW conversions only;
    \item Spectral swaps, followed by MSW conversions;
    \item Spectral swaps, FFCs and MSW conversions, in this order;
\end{itemize}
for both orderings (NO vs.~IO).
We treat the different mechanisms sequentially, taking into account the ubiquitousness of MSW conversions, even in the absence of collective effects. Moreover, in line with the discussion in~\cref{sec:flacon}, we apply the effect of FFCs -- whose occurrence depends on the angular distributions and may be absent in certain configurations -- after that of slow instabilities, which can develop already at the early stages of the evolution~\cite{Fiorillo:2025gkw}.
Together with the (ordering-independent) ``unoscillated'' case, where no flavour conversions occur, this corresponds to a total of 7 possible frameworks for the interpretation of data from a future SN event.

\subsection{Supernova flux parameters}

\begin{figure}[p]
\vskip -2mm
\hspace{-2.5cm}
    \includegraphics[width=1.3\textwidth]{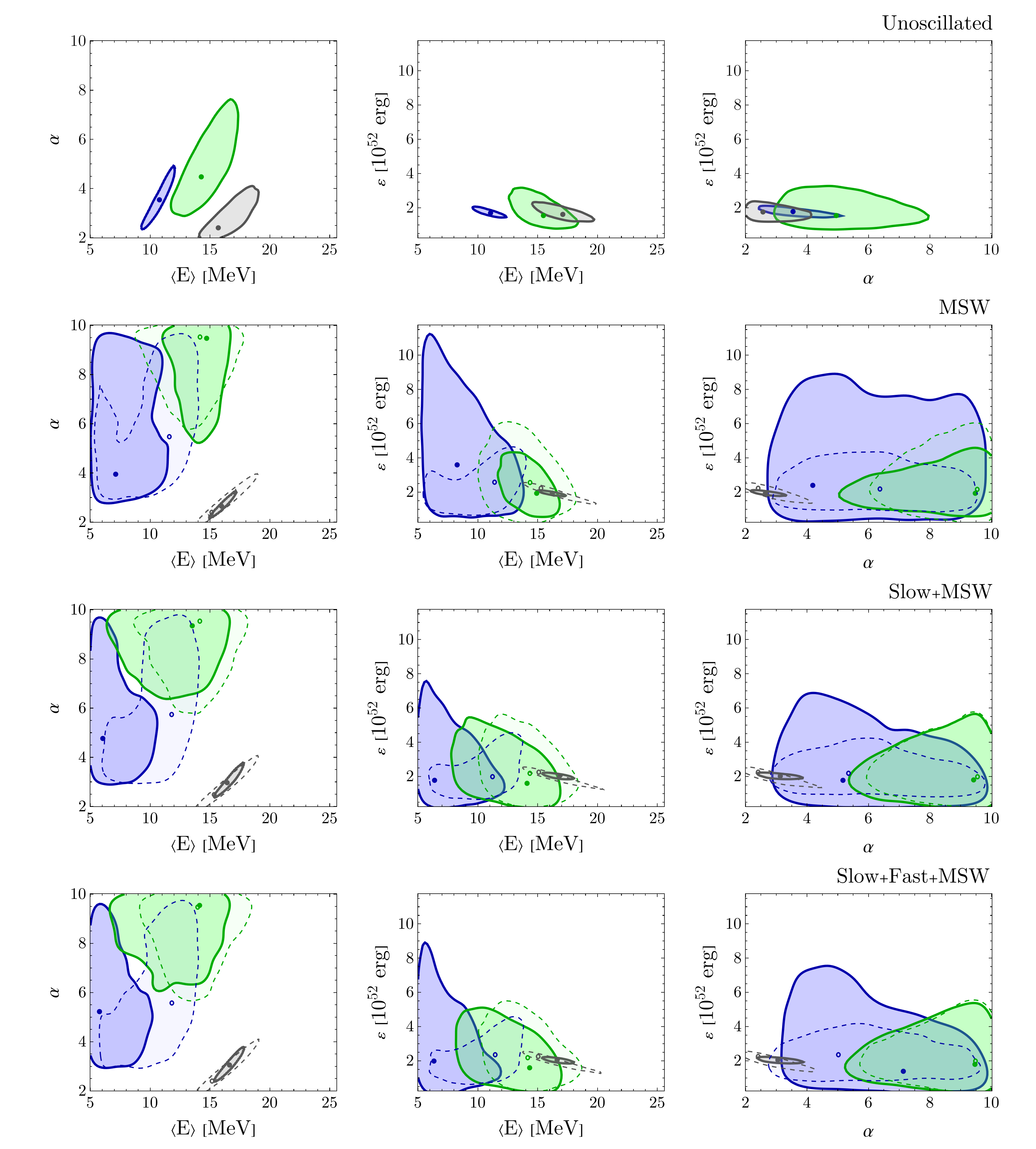}
\vskip 2mm
    \caption{
    Credible regions (1$\sigma$) for neutrino spectral parameters inferred for benchmark \A,
    in the $\langle E\rangle_i-\alpha_i$, $\langle E\rangle_i-\epsilon_i$, and $\alpha_i-\epsilon_i$ planes. Each row corresponds to a different flavour-conversion scenario. Markers indicate the 2D posterior modes. Solid contours and filled markers correspond to NO, while dashed contours and open markers pertain to IO. Colours identify the neutrino species $i$, with $\nu_e$ in blue, $\bar\nu_e$ in green, and $\nu_x$ in grey.
    }
    \label{fig:A1s}\vspace{-82pt}
\end{figure}

In this section, we investigate the capability of DUNE to reconstruct the supernova neutrino flux parameters across the different flavour conversion scenarios described in the previous section. Our goal is to quantify the impact of collective oscillations not only on the neutrino fluences, but, more importantly, on the reconstructed spectra observable at DUNE.
To this end, we have interfaced the GLoBES event rate generator implementing the various conversion scenarios with the nested sampling algorithm MultiNest~\cite{Feroz:2013hea}.
MultiNest provides a Bayesian inference framework that efficiently explores high-dimensional parameter spaces by evolving a set of live points distributed over the prior volume. At each iteration, the point with the lowest likelihood is replaced by a new sample drawn with higher likelihood, progressively concentrating on the regions most favoured by the data. In addition, MultiNest provides posterior distributions and Bayesian evidence estimates, enabling a statistically-consistent determination of credible regions and Bayes factors for model comparison, through repeated calls to the
conversion probability and event rate subroutines. 
In our setup, we compute the expected neutrino spectra using GLoBES, taking as true values the benchmark parameters summarized in~\cref{tab:benchmarks}, and scan the parameter space as described in~\cref{sec:benchmarks}. We then derive
the $1\sigma$ credible regions for the spectral parameters in the two-dimensional planes involving $\alpha_i$, $\langle E_i\rangle$, and $\epsilon_i$, for each species $i$,
marginalizing over all parameters not shown. 
The results for benchmark~\A are collected in~\cref{fig:A1s}.

\begin{figure}[p]
\vskip -2mm
\hspace{-2.5cm}
    \includegraphics[width=1.3\textwidth]{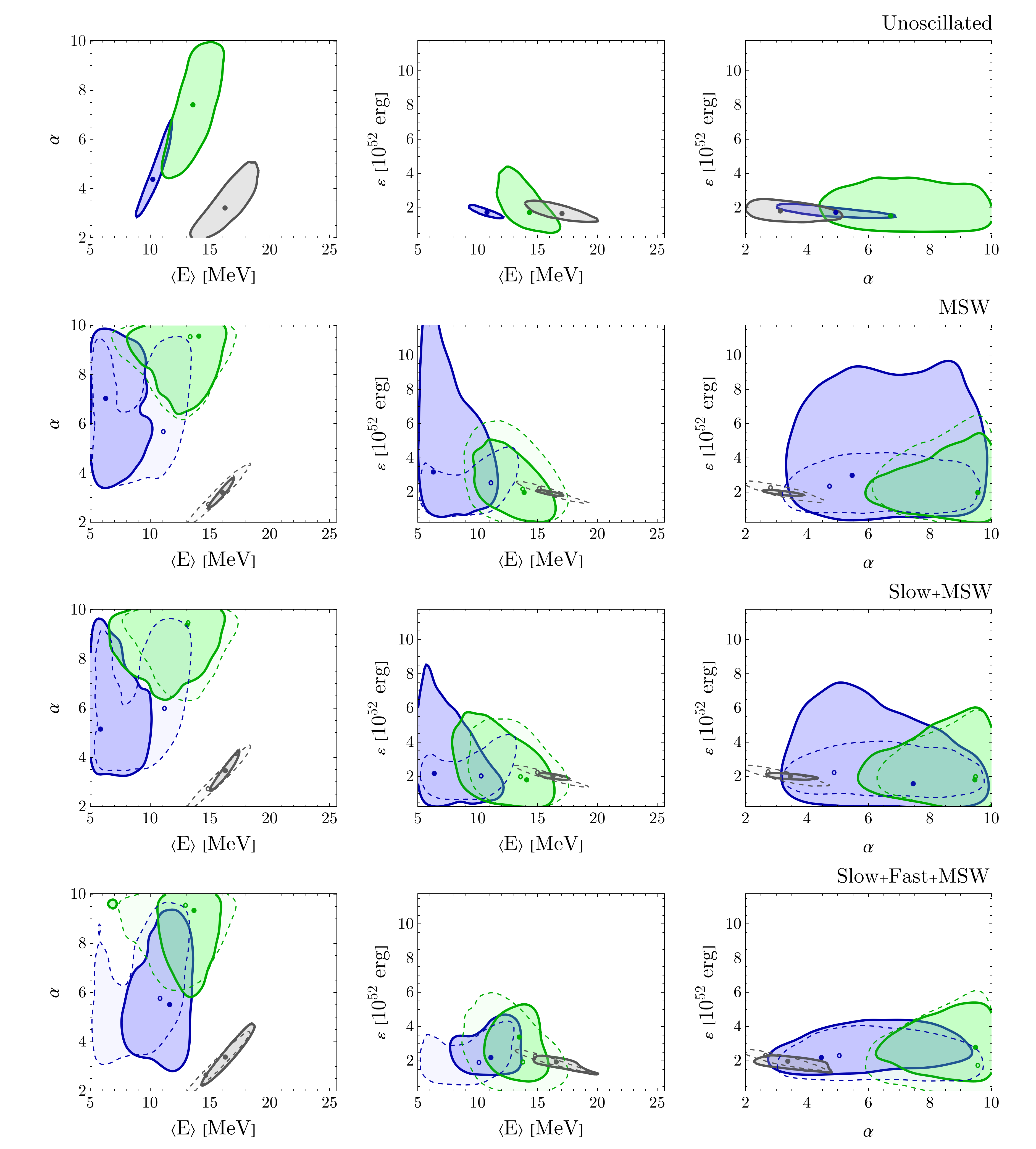}
\vskip 2mm
    \caption{The same as~\cref{fig:A1s}, for benchmark \B.}
    \label{fig:B1s}\vspace{-34pt}
\end{figure}

In the first row of~\cref{fig:A1s}, the unoscillated case is presented, with blue, green, and grey regions depicting the credible regions for $\nu_e$, $\bar\nu_e$, and $\nu_x$ parameters, respectively. The coloured markers inside these regions indicate the 2D posterior modes, which do not necessarily coincide with the true values, due to the nature of the inference procedure. As discussed in~\cref{sec:DUNE}, the $\nu_e$ CC channel yields the largest number of expected events; consequently, $\alpha_{\nu_e}$, $\varepsilon_{\nu_e}$, and $\langle E_{\nu_e}\rangle$ are reconstructed with high precision in the unoscillated case.
The $\nu_x$ parameters are slightly less constrained. Namely, the pinching $\alpha_{\nu_x}$ is expected to be harder to determine, since most $\nu_x$ events are detected via the NC channel and collectively reconstructed at low energies. One finds that the lower bound on $\alpha_{\nu_x}$ coincides with the lower limit of the prior range. Nevertheless, the prior inequality forces it to be the smallest among the three $\alpha_i$, allowing an upper bound to be established.
By contrast, the $\bar\nu_e$ parameters are poorly constrained, as the $\bar\nu_e$ CC channel has both the smallest cross section and the highest energy threshold ($E\gtrsim16$ MeV). The best-measured quantity in this sector is $\langle E_{\bar\nu_e}\rangle$, whose determination benefits from the inequality $\langle E_{\nu_e}\rangle \leq \langle E_{\bar\nu_e}\rangle \leq \langle E_{\nu_x}\rangle$, which restricts it to a relatively narrow range.

The inclusion of MSW conversions drastically modifies these results, as shown in the second row of~\cref{fig:A1s}, where solid and dashed contours correspond to NO and IO, respectively.
In the NO case, the vast majority of electron neutrinos are converted into other flavours upon exiting the supernova medium (recall~\cref{eq:MSWfluences,eq:MSWprobs}). As a result, the initial $\nu_e$ fluence parameters can only be probed via NC interactions of the converted $\nu_x$ component. Since the NC cross section in DUNE has a high energy threshold ($E \gtrsim 10$~MeV), only a small fraction of the originally-electron neutrinos are detected, leading to a systematic underestimation of $\langle E_{\nu_e}\rangle$ -- for which an upper limit is still set due to the imposed prior inequalities -- and allowing only a lower bound to be placed on $\alpha_{\nu_e}$, enforced by the prior constraint $\alpha_{\nu_e} \geq \alpha_{\nu_x}$. Conversely, since most $\nu_x$ are converted into $\nu_e$, the $\nu_x$ fluence parameters are now reconstructed with excellent precision. For $\bar\nu_e$, the conversion is less complete, though the fraction converting into $\bar\nu_x$ still significantly weakens the determination of the corresponding pinching parameter.
In the IO case, the roles of neutrinos and antineutrinos are interchanged: approximately 70\% of $\nu_e$ are converted into $\nu_x$, yielding slightly improved constraints on $\nu_e$ parameters and slightly weaker ones on those of $\nu_x$, when compared to NO. Meanwhile, nearly all $\bar\nu_e$ are converted into $\bar\nu_x$, degrading the corresponding parameter bounds, even though the prior inequalities still provide partial anchoring.
This comparison between the unoscillated and standard MSW scenarios illustrates clearly how critical it is to adopt the correct flavour-conversion model when analysing SN neutrino data.

In the lower two rows of~\cref{fig:A1s}, we show the results for the slow conversions + MSW and slow + fast conversions + MSW scenarios. The interpretation of these cases is more involved, but the results remain dominated by the subsequent MSW conversions, which tend to improve the constraints on $\nu_x$ parameters while significantly degrading those on $\nu_e$. Notably, the presence of spectral swaps induced by slow instabilities can help tighten the bounds on certain parameters, such as the luminosity $\varepsilon_{\nu_e}$.

For the sake of comparison,~\cref{fig:B1s} shows the results for benchmark~\B, in which the three initial fluences are assigned distinct pinching parameters (cf.~\cref{tab:benchmarks}). Without repeating a detailed analysis for each case -- which would display the same qualitative features as those identified for benchmark~{\A} -- we highlight two notable differences.
Firstly, the higher pinching of the initial $\nu_e$ fluence ($\alpha_{\nu_e} = 3.5$, compared to 2.5 in benchmark~\A) leads to a slightly improved reconstruction of the $\nu_e$ parameters under MSW conversions in NO. This can be understood from the fact that, for a pinched flux, the peak of the energy distribution is located at
\begin{equation}
    E_\mathrm{peak}=\frac{\alpha}{\alpha+1}\langle E\rangle \, ,
\end{equation}
so that a larger pinching parameter shifts the peak to higher energies, increasing the fraction of electron-flavour neutrinos with energies above the NC detection threshold. Additionally, in the scenario including fast conversions, the equalization of fluences prior to MSW effects leads to an improved determination of $\langle E_{\nu_e}\rangle$ and $\varepsilon_{\nu_e}$ in NO, with a relatively worse determination of the original $\nu_x$ parameters as a trade-off.

\afterpage{
\begin{landscape}
%%%%%%%%%%%%%%%%%%%%%%
\begin{table}[p]
\thisfloatpagestyle{empty}
\hspace{-4cm} \small
\renewcommand{\arraystretch}{1.2}
\begin{tabular}{llccccccccc}
\toprule
\multicolumn{2}{c}{\multirow{2}{*}{Benchmark \A}}
& \multicolumn{3}{c}{$\epsilon$ ($10^{52}$ erg)} 
& \multicolumn{3}{c}{$\alpha$}
& \multicolumn{3}{c}{$\langle E \rangle$ (MeV)}
\\
& & $\epsilon_{\nu_e}$ & $\epsilon_{\bar \nu_e}$ & $\epsilon_{\nu_x}$ 
& $\alpha_{\nu_e}$ & $\alpha_{\bar \nu_e}$ & $\alpha_{\nu_x}$ 
& $\langle E_{\nu_e} \rangle$ & $\langle E_{\bar\nu_e} \rangle$ & $\langle E_{\nu_x} \rangle$ 
\\
\midrule
\multicolumn{2}{c}{Unoscillated} 
&  $-0.30\,(-0.54)$ &  $+0.18\,(+0.08)$ &  $-0.32\,(-0.39)$ &  $+2.16\,(+0.94)$ &  $+4.40\,(+1.97)$ &  $+1.05\,(+0.52)$ &  $+2.01\,(+1.23)$ &  $+3.45\,(+1.48)$ &  $+2.38\,(+0.89)$ \\
\midrule
\multirow{3}{*}{NO} & MSW & $+3.60\,(+0.82)$ &  $+0.96\,(+0.28)$ &  $-0.10\,(-0.34)$ &  $+4.32\,(+2.12)$ &  $+6.16\,(+3.81)$ &  $+0.46\,(+0.29)$ &  $-0.72\,(-0.25)$ &  $+2.25\,(+0.82)$ &  $+0.99\,(+0.47)$ \\
& Slow+MSW &  $+2.38\,(+0.56)$ &  $+1.54\,(+0.40)$ &  $-0.02\,(-0.07)$ &  $+4.17\,(+2.11)$ &  $+6.27\,(+4.38)$ &  $+0.86\,(+0.52)$ &  $-1.51\,(-0.61)$ &  $+0.71\,(+0.21)$ &  $+1.55\,(+0.78)$ \\
& Slow+Fast+MSW &  $+2.43\,(+0.58)$ &  $+1.41\,(+0.39)$ &  $-0.03\,(-0.08)$ &  $+4.36\,(+2.22)$ &  $+6.32\,(+4.34)$ &  $+0.90\,(+0.53)$ &  $-1.62\,(-0.63)$ &  $+0.99\,(+0.29)$ &  $+1.63\,(+0.79)$ \\
\midrule
\multirow{3}{*}{IO} & MSW  & $+1.19\,(+0.43)$ &  $+1.48\,(+0.44)$ &  $-0.24\,(-0.47)$ &  $+4.52\,(+2.26)$ &  $+6.40\,(+4.64)$ &  $+1.04\,(+0.56)$ &  $+0.88\,(+0.29)$ &  $+2.16\,(+0.62)$ &  $+2.26\,(+0.78)$ \\
& Slow+MSW & $+0.92\,(+0.35)$ &  $+1.26\,(+0.36)$ &  $-0.23\,(-0.45)$ &  $+4.53\,(+2.25)$ &  $+6.38\,(+4.38)$ &  $+1.04\,(+0.55)$ &  $+1.00\,(+0.34)$ &  $+2.18\,(+0.63)$ &  $+2.25\,(+0.77)$ \\
& Slow+Fast+MSW &  $+0.96\,(+0.35)$ &  $+1.17\,(+0.36)$ &  $-0.24\,(-0.48)$ &  $+4.49\,(+2.22)$ &  $+6.40\,(+4.31)$ &  $+1.12\,(+0.58)$ &  $+0.97\,(+0.33)$ &  $+2.20\,(+0.63)$ &  $+2.39\,(+0.79)$ \\

\bottomrule
\end{tabular}\captionsetup{width=23.5cm}
\caption{
Distance between the 1D posterior median and the true parameter value, for benchmark \A, for each flavour-conversion scenario and spectral parameter.
Values in parentheses denote the corresponding pull, i.e.~the reported distance in units of the posterior standard deviation.
}
\label{tab:distA}
\end{table}
%%%%%%%%%%%%%%%%%%%%%%

%%%%%%%%%%%%%%%%%%%%%%
\begin{table}[p]
\thisfloatpagestyle{empty}
\hspace{-4cm} \small
\renewcommand{\arraystretch}{1.2}
\begin{tabular}{llccccccccc}
\toprule
\multicolumn{2}{c}{\multirow{2}{*}{Benchmark \B}}
& \multicolumn{3}{c}{$\epsilon$ ($10^{52}$ erg)} 
& \multicolumn{3}{c}{$\alpha$}
& \multicolumn{3}{c}{$\langle E \rangle$ (MeV)}
\\
& & $\epsilon_{\nu_e}$ & $\epsilon_{\bar \nu_e}$ & $\epsilon_{\nu_x}$ 
& $\alpha_{\nu_e}$ & $\alpha_{\bar \nu_e}$ & $\alpha_{\nu_x}$ 
& $\langle E_{\nu_e} \rangle$ & $\langle E_{\bar\nu_e} \rangle$ & $\langle E_{\nu_x} \rangle$ 
\\
\midrule
\multicolumn{2}{c}{Unoscillated} &  $-0.22\,(-0.26)$ &  $+0.55\,(+0.20)$ &  $-0.23\,(-0.22)$ &  $+2.24\,(+1.06)$ &  $+4.08\,(+2.33)$ &  $+1.15\,(+0.58)$ &  $+1.34\,(+0.96)$ &  $+1.97\,(+0.89)$ &  $+1.75\,(+0.66)$ \\
\midrule
\multirow{3}{*}{NO} & MSW &  $+4.10\,(+0.92)$ &  $+1.41\,(+0.36)$ &  $-0.07\,(-0.24)$ &  $+3.75\,(+2.03)$ &  $+4.76\,(+3.64)$ &  $+0.44\,(+0.27)$ &  $-1.36\,(-0.50)$ &  $+1.07\,(+0.38)$ &  $+0.76\,(+0.40)$ \\
& Slow+MSW  &  $+2.68\,(+0.63)$ &  $+1.79\,(+0.45)$ &  $-0.03\,(-0.10)$ &  $+3.62\,(+1.89)$ &  $+4.72\,(+3.48)$ &  $+0.75\,(+0.44)$ &  $-1.61\,(-0.68)$ &  $+0.27\,(+0.09)$ &  $+1.14\,(+0.59)$ \\
& Slow+Fast+MSW &  $+1.69\,(+0.42)$ &  $+1.48\,(+0.41)$ &  $-0.12\,(-0.30)$ &  $+3.51\,(+1.82)$ &  $+4.72\,(+3.67)$ &  $+1.04\,(+0.66)$ &  $+0.13\,(+0.05)$ &  $+1.05\,(+0.34)$ &  $+1.96\,(+0.99)$ \\
\midrule
\multirow{3}{*}{IO} & MSW & $+1.25\,(+0.41)$ &  $+1.63\,(+0.46)$ &  $-0.17\,(-0.32)$ &  $+3.75\,(+1.94)$ &  $+4.79\,(+3.50)$ &  $+0.85\,(+0.44)$ &  $+0.05\,(+0.02)$ &  $+1.29\,(+0.37)$ &  $+1.56\,(+0.56)$ \\
& Slow+MSW & $+1.03\,(+0.37)$ &  $+1.35\,(+0.38)$ &  $-0.16\,(-0.30)$ &  $+3.92\,(+2.03)$ &  $+4.77\,(+3.52)$ &  $+0.94\,(+0.47)$ &  $+0.21\,(+0.07)$ &  $+1.39\,(+0.40)$ &  $+1.65\,(+0.57)$ \\
& Slow+Fast+MSW &  $+0.99\,(+0.37)$ &  $+1.37\,(+0.39)$ &  $-0.15\,(-0.28)$ &  $+3.80\,(+1.92)$ &  $+4.75\,(+3.37)$ &  $+0.85\,(+0.42)$ &  $+0.20\,(+0.07)$ &  $+1.27\,(+0.35)$ &  $+1.55\,(+0.52)$ \\

\bottomrule
\end{tabular}
    \caption{The same as~\cref{tab:distA}, for benchmark \B.}
\label{tab:distB}
\end{table}
%%%%%%%%%%%%%%%%%%%%%%
%
\end{landscape}
}

\vskip 2mm
\Cref{tab:distA,tab:distB} summarize the pulls of the one-dimensional posterior medians with respect to the benchmark true values, for benchmarks~\A and~\B respectively. We adopt posterior medians rather than means to better capture asymmetric posteriors, with all statistical quantities derived directly from the MultiNest samples and uncertainties expressed as posterior standard deviations. We caution that posterior medians need not coincide with the posterior mode or with the true parameter values, and their interpretation should be guided by the full credible regions shown in~\cref{fig:A1s,fig:B1s}.

For benchmark~\A, in the unoscillated case, nearly all true parameter values lie within one posterior standard deviation of the median, with three notable exceptions. The parameters $\alpha_{\bar\nu_e}$ and $\langle E_{\bar\nu_e}\rangle$ are poorly recovered, a direct consequence of the limited statistics available in the $\bar\nu_e$ CC channel, which, as discussed above, has both the smallest cross section and the highest energy threshold in DUNE. The third outlier, $\langle E_{\nu_e}\rangle$, is displaced from its true value by approximately $1.2\sigma$, a subtler effect arising from the positive correlation between the pinching parameter $\alpha$ and the mean energy $\langle E\rangle$ in the pinched-thermal parameterization (visible in the top panels of~\cref{fig:A1s,fig:B1s}): the prior constraint $\alpha_{\nu_e} \geq \alpha_{\nu_x}$ pushes the posterior of $\alpha_{\nu_e}$ toward higher values, which in turn pulls the median of $\langle E_{\nu_e}\rangle$ upward, and the high precision of this channel amplifies the resulting tension.

In all conversion scenarios within benchmark~\A, the luminosity parameters $\epsilon_i$ remain the most reliably reconstructed quantities, with posterior medians generally within one standard deviation of the true values. The notable exception is $\epsilon_{\nu_e}$ under NO, where the near-complete conversion of $\nu_e$ into other flavours (cf.~\cref{eq:MSWfluences,eq:MSWprobs}) drives a relatively large absolute shift of the posterior median toward higher values. The pinching parameters of the electron neutrino and antineutrino species are the most adversely affected by flavour conversions: $\alpha_{\nu_e}$ and $\alpha_{\bar\nu_e}$ are systematically overestimated, with the true values lying at distances ranging from approximately $2\sigma$ (for $\alpha_{\nu_e}$) to $4.5\sigma$ (for $\alpha_{\bar\nu_e}$)
for the posterior medians.
Conversely, as expected from the discussion of~\cref{fig:A1s}, the conversion of $\nuxbp$ into $\nuebp$ via MSW effects enables a satisfactory reconstruction of the non-electron-species pinching parameters across all scenarios.
The mean energies $\langle E_i \rangle$ are generally recovered within one standard deviation, though with a tendency toward systematic overestimation. 
The exception is $\langle E_{\nu_e} \rangle$, 
for which the sign of the residual deviation reflects the mass ordering: deviations are negative in NO and positive in IO.
No significant qualitative differences are found among the three collective oscillation mechanisms considered, pointing to the MSW layer as the dominant driver of parameter-reconstruction quality.

For benchmark~\B, the unoscillated case yields a slightly improved recovery of the mean energies, due to the larger spectral hierarchy among flavours. However, the reconstruction of the pinching parameters is somewhat degraded relative to benchmark~\A: since the three $\alpha_i$ values are now distinct and more widely separated, the assumed prior inequality constraints are less effective at anchoring the individual posteriors. In particular, $\alpha_{\bar\nu_e}$, which is unconstrained from above, is substantially overestimated (${\sim}2.3\sigma$). In flavour conversion cases, the posterior medians are overall slightly closer to the true values, driven primarily by the richer spectral structure that allows better separation of the individual fluences.

\vskip 2mm
Taken together, the results for both benchmarks lead to a consistent conclusion: while collective oscillations induce significant modifications to the observed spectra at DUNE, correctly accounting for them in the likelihood does not substantially degrade the experiment's ability to constrain the source parameters, relative to the unoscillated baseline. The most significant deterioration in reconstruction quality is instead attributable to the MSW effect, which considerably reshuffles the initial flavour composition of the supernova neutrino burst and thereby suppresses the sensitivity to the original electron-flavour fluences.

\subsection{Distinguishing physical scenarios}

%%%%%%%%%%%%%%%%%%%%%%
\definecolor{weak}{RGB}{245,250,245}
\definecolor{moderate}{RGB}{230,245,230}
\definecolor{strong}{RGB}{210,240,210}
\definecolor{moderateagainst}{RGB}{245,220,220}
\definecolor{inconclusive}{gray}{0.99}

\newcommand\inc{\cellcolor{inconclusive}}
\newcommand\ag{\cellcolor{moderateagainst}}
\newcommand\weak{\cellcolor{weak}}
\newcommand\md{\cellcolor{moderate}}
\newcommand\str{\cellcolor{strong}}

\begin{table}[p]
\centering
\renewcommand{\arraystretch}{1.2}
\begin{tabular}{llccccccc}
\toprule
\multicolumn{2}{c}{\multirow{2}{*}{Benchmark \A}}
& \multirow{2}{*}{Unosc.}
& \multicolumn{3}{c}{NO} 
& \multicolumn{3}{c}{IO}
\\
& & 
& M & S+M & S+F+M
& M & S+M & S+F+M
\\
\midrule
\multicolumn{2}{c}{Unosc.} & 
{\inc}0.00 & 
{\str}42.71 & {\str}22.85 & {\str}20.86 & 
{\str}25.15 & {\str}24.46 & {\str}24.49
\\
\midrule
\multirow{3}{*}{NO} & M &
{\str}4.63 & 
{\inc}0.00 & {\str}5.09 & {\str}4.16 & 
{\str}2.81 & {\str}2.73 & {\str}2.72
\\
& S+M  &  
{\str}6.58 & 
{\str}6.43 & {\inc}0.00 & {\inc}0.07 &
{\md}1.28 & {\md}1.18 & {\md}1.19
\\
& S+F+M & 
{\str}6.66 &
{\str}6.35 & {\inc}0.02 & {\inc}0.00 &
{\md}1.22 & {\md}1.11 & {\md}1.09
\\
\midrule
\multirow{3}{*}{IO} & M &  
{\str}5.71 & 
{\str}3.82 & {\str}2.28 & {\md}1.41 &
{\inc}0.00 & {\inc}0.23 & {\inc}0.24
\\
& S+M  & 
{\str}5.60 & 
{\str}3.71 & {\str}2.10 & {\md}1.17 &
{\inc}$-$0.19 & {\inc}0.00 & {\inc}$-$0.02
\\
& S+F+M & 
{\str}5.54 & 
{\str}3.70 & {\str}2.12 & {\md}1.13 &
{\inc}$-$0.14 & {\inc}0.03 & {\inc}0.00
\\
\bottomrule
\end{tabular}
\caption{
Logarithm (base 10) of the Bayes factors for benchmark \A, comparing different flavour-conversion scenarios.
Rows correspond to the true (data-generating) scenarios, while columns denote the tested hypotheses (M = MSW, S = slow, F = fast).
Positive values indicate evidence against the tested scenario relative to the true one. Cell colours encode the strength of discrimination, with greener shades corresponding to increasing discriminatory power and grey indicating inconclusive results.
}
\label{tab:bayesA}
\end{table}
%%%%%%%%%%%%%%%%%%%%%%

%%%%%%%%%%%%%%%%%%%%%%
\begin{table}[p]
\centering
\renewcommand{\arraystretch}{1.2}
\begin{tabular}{llccccccc}
\toprule
\multicolumn{2}{c}{\multirow{2}{*}{Benchmark \B}}
& \multirow{2}{*}{Unosc.}
& \multicolumn{3}{c}{NO} 
& \multicolumn{3}{c}{IO}
\\
& & 
& M & S+M & S+F+M
& M & S+M & S+F+M
\\
\midrule
\multicolumn{2}{c}{Unosc.} & 
{\inc}0.00 &
{\str}46.05 & {\str}27.49 & {\str}27.50 &
{\str}30.90 & {\str}29.87 & {\str}30.42
\\
\midrule
\multirow{3}{*}{NO} & M & 
{\str}4.69 &
{\inc}0.00 & {\str}5.29 & {\str}4.75 &
{\str}3.46 & {\str}3.19 & {\str}3.23
\\
& S+M  & 
{\str}5.87 &
{\str}5.57 & {\inc}0.00 & {\inc}-0.07 &
{\md}1.14 & {\md}1.01 & {\md}1.00
\\
& S+F+M & 
{\str}4.25 &
{\str}2.13 & {\weak}0.98 & {\inc}0.00 &
{\ag}$-$0.58 & {\ag}$-$0.55 & {\ag}$-$0.62
\\
\midrule
\multirow{3}{*}{IO} & M & 
{\str}5.06 & 
{\str}3.24 & {\str}2.34 & {\md}1.47 &
{\inc}0.00 & {\inc}0.06 & {\inc}0.18
\\
& S+M  &  
{\str}5.13 & 
{\str}3.22 & {\str}2.21 & {\md}1.31 &
{\inc}$-$0.16 & {\inc}0.00 & {\inc}$-$0.03
\\
& S+F+M & 
{\str}5.24 & 
{\str}3.30 & {\str}2.28 & {\md}1.25 &
{\inc}$-$0.13 & {\inc}$-$0.07 & {\inc}0.00
\\
\bottomrule
\end{tabular}
\caption{The same as~\cref{tab:bayesA}, for benchmark \B, with pale-red cells indicating moderate evidence in favour of the tested scenario.}
\label{tab:bayesB}
\end{table}
%%%%%%%%%%%%%%%%%%%%%%

In this section, we assess the performance of DUNE in discriminating among the various flavour-conversion scenarios introduced above, following an approach similar to the one proposed in~\cite{Abbar:2024nhz} in the context of water-Cherenkov detectors. To this end, we simulate data under each of the considered hypotheses -- unoscillated, MSW only (M), slow conversions + MSW (S+M), and slow + fast conversions + MSW (S+F+M), for both mass orderings -- and perform Bayesian inference under each of the remaining hypotheses, adopting a model comparison approach based on the evidence computed by MultiNest. The Bayes factor is defined as the ratio of the Bayesian evidences of the true and tested scenarios,
\begin{equation}
    B = \frac{\pazocal{Z}_\mathrm{true}}{\pazocal{Z}_\mathrm{test}} \, ,
\end{equation}
where the evidence corresponds to the likelihood marginalized over the full parameter space with prior weighting,
\begin{equation}
    \pazocal{Z}_m = \int \pazocal{L}(\theta_m)\,\pi(\theta_m)\, \mathrm{d}\theta_m \, ,
\end{equation}
with $\pazocal{L}(\theta_m)$ and $\pi(\theta_m)$ denoting the likelihood function and prior for the parameters $\theta_m$ of model $m$, respectively. This formulation ensures that the comparison penalizes unnecessary model complexity, reflecting not only how well the model can explain the data but also the volume of parameter space consistent with it.
We present $\log_{10} B$ in~\cref{tab:bayesA,tab:bayesB}, adopting the convention that positive values indicate evidence against the tested scenario (columns) relative to the true one (rows). We interpret the results in line with Jeffreys' scale~\cite{Jeffreys:1998}: $\log_{10}B > 2$ constitutes decisive evidence, $1-2$ strong evidence, $0.5-1$ moderate evidence, $|{\log_{10}B}| < 0.5$ inconclusive, and $\log_{10}B < -0.5$ moderate evidence in favour of the tested scenario.

For benchmark~\A, the unoscillated scenario is decisively distinguished from all conversion ones: $\log_{10}B > 20$ when data are generated under the unoscillated hypothesis, and $\log_{10}B > 4$ in the converse case. This constitutes the first robust conclusion of our analysis: the unoscillated case can always be distinguished from other conversion scenarios, provided MSW conversions are at play. Among the conversion scenarios in NO, the MSW-only hypothesis (M) is decisively distinguishable from all others, both as the true and as the tested model. The situation is markedly different in IO: the MSW-only scenario (M) is indistinguishable from the other IO conversion cases ($|\log_{10}B| < 0.3$), and strongly but not, overall, decisively distinguishable when compared with NO scenarios including collective effects (S+M and S+F+M). This leads to the conclusion that under the IO hypothesis, different conversion mechanisms can produce observationally similar spectra at DUNE, rendering the identification of the correct conversion scenario significantly more challenging. 
On the contrary, the MSW-only (M) scenario in NO remains distinctive with respect to all other considered conversion scenarios -- all three IO cases (M, S+M, S+F+M) and the two NO scenarios including collective effects (S+M, S+F+M) -- precisely because of the near-complete conversion of electron-flavour neutrinos into non-electron flavours, which produces a qualitatively unique signature.

Turning to the scenarios including collective effects, we find that the S+M and S+F+M schemes cannot be mutually distinguished when the mass ordering is the same in the true and tested hypotheses. This demonstrates that DUNE lacks the sensitivity to resolve the subtle spectral imprints of fast flavour conversions, at least in the benchmark configurations considered here, where full flavour equilibration is not necessarily reached after FFCs. When data generation involves collective effects in NO and the test hypothesis corresponds to IO with collective effects, moderate to strong discrimination is still achievable ($\log_{10}B \sim 1$). Conversely, when data is generated with collective effects in IO and the true hypothesis is compared with a test hypothesis based on NO with collective effects, a stronger discrimination is possible. Nevertheless, the presence of fast conversions reduces the discriminatory power ($\log_{10}B \sim 2$ for S+M and $\log_{10}B \sim 1$ for S+F+M as test hypotheses). Indeed, the milder spectral modifications induced by MSW in IO can be partially mimicked by MSW in NO when attenuated by fast conversions driving the fluences toward flavour equilibration.

For benchmark~\B, all the above conclusions remain qualitatively valid. The most notable difference arises when data are generated under S+F+M in NO: in this case, the discriminatory power against the unoscillated ($\log_{10}B: 6.66 \to 4.25$) and MSW-only ($\log_{10}B: 6.35 \to 2.13$) scenarios in NO is reduced relative to benchmark~\A, while it increases against S+M ($\log_{10}B: 0.02 \to 0.98$). This pattern suggests that, in benchmark~\B, fast conversions suppress the spectral imprints of slow conversions more effectively, bringing the resulting spectra closer to those expected in the absence of collective effects. Furthermore, the S+F+M scenario in NO is found to be moderately better described by IO hypotheses (M, S+M, or S+F+M in IO, with $\log_{10}B \lesssim -0.5$), indicating that collective effects can partially compensate the drastic spectral reshuffling induced by MSW in NO, rendering the resulting spectra more similar to those characteristic of IO, in a sufficiently large volume of parameter space. These last findings, resulting from the comparison of different benchmarks, carry an important practical implication: the reconstruction and interpretation of SN neutrino data depend not only on the assumed flavour conversion model, but also on the spectral parameters characterizing the initial neutrino fluences; this ambiguity becomes particularly important when multiple conversion mechanisms operate simultaneously.

%%%%%%%%%%%%%%%%%%%%%%%%%%%%%%%%%%%%%%%%%%%%%%
\section{Summary and conclusions}
\label{sec:summary}
%%%%%%%%%%%%%%%%%%%%%%%%%%%%%%%%%%%%%%%%%%%%%%

Collective flavour conversions can strongly affect the composition of the neutrino signal coming from core-collapse supernovae.
Being based on standard physical processes, they must be taken into account in any realistic interpretation of the data, 
whether in the case of a single explosion or in the analysis of the diffuse supernova neutrino background.
However, the connection between simulations
of flavour conversions during the supernova evolution 
and of flavour-dependent event rates
is still tenuous and deserves further attention.
With this in mind, we adopt a phenomenological approach to the simulation of terrestrial event rates, using established analytical prescriptions to incorporate the effects of slow and fast flavour instabilities on SN neutrino fluxes. 

\vskip 2mm
Employing the aforementioned approach, described in~\cref{sec:SNnus}, we have examined in detail the extent to which DUNE will be able to constrain the parameters characterizing the original SN neutrino fluxes, under different assumptions regarding the neutrino mass ordering (NO vs.~IO) and the active flavour conversion mechanisms -- to wit, spectral swaps induced by slow instabilities, fast flavour conversions, and MSW conversions -- which were applied sequentially.
The pair of considered benchmarks is described in~\cref{tab:benchmarks}.
The prospective capabilities of DUNE make it a key future contributor to a global network of SN neutrino observatories. 
While able to detect neutrinos and antineutrinos of all flavours, DUNE will be particularly sensitive to the electron-neutrino supernova flux (cf.~\cref{sec:DUNE}), providing valuable data towards testing different models of SN emission and flavour conversion.

Our results on Bayesian parameter estimation are summarized in~\cref{fig:A1s,fig:B1s,tab:distA,tab:distB}.  
We find that flux parameters can be inferred with varying degrees of accuracy and precision, depending on the active flavour conversion mechanism(s).
Naturally, the parameters whose estimation relies on the electron-neutrino channel are determined with better precision.
In particular, MSW conversions deplete a significant portion of the original  $\nu_e$ fluence, considerably suppressing the sensitivity to the corresponding parameters, while enabling instead a satisfactory reconstruction of the original non-electron-flavour spectrum, part of which has been converted into a discernible $\nu_e$ flux.  
While the presence of collective effects may bring about particular spectral signatures, it does not dramatically degrade or enhance the ability of DUNE to constrain the luminosity, pinching and average-energy parameters,
provided the correct model is used to interpret the data.
We also stress the importance of physics-informed prior inequalities for parameter estimation. These encode SN simulation expectations directly into the sampling procedure and contribute to bounding quantities that would not be sufficiently constrained by the prospective data.

It is crucial to adopt the correct flavour-conversion model when using SN neutrino data to extract physical information.
We have therefore quantified DUNE's ability to discriminate between different flavour conversion scenarios.
Our results on model discrimination using relative Bayesian evidences are summarized in~\cref{tab:bayesA,tab:bayesB}.
We find that the no-conversions case can always be distinguished from other scenarios where MSW conversions are at play.
Moreover, for the considered benchmarks, one can almost always obtain a strong-evidence identification of the correct neutrino mass ordering across conversion scenarios.
In contrast, while an MSW-only conversion scenario can be decisively distinguished from scenarios involving collective oscillations in the NO case, no such discrimination seems possible within IO.
Finally, we find that, for a given mass ordering, DUNE is not able to ascertain the presence or absence of fast flavour conversions in a framework where both MSW and slow conversions are also active.

\vskip 2mm

Our work showcases the challenges in SN neutrino data interpretation and the crucial role played by the knowledge of the neutrino phase-space distribution.
Indeed, this interpretation depends not only on the assumed flavour conversion model, but also on the spectral parameters characterizing the initial fluences.
While our results are quantitative, they are not definitive, and several avenues for refinement remain, e.g.~tracking the spectral evolution over time.
Future work is also required to strengthen the bridge between SN and detector simulations, in anticipation of the next Galactic event.

%%%%%%%%%%%%%%%%%%%%%%%%%%%%%%%%%%%%%%%%%%%%%%
\footnotesize
%%%%%%%%%%%%%%%%%%%%%%%
\bibliographystyle{JHEPwithnote.bst}
\bibliography{bibliography}
%%%%%%%%%%%%%%%%%%%%%%%

\end{document}